\documentclass[preprint,nofootinbib,floats,tightenlines,amsfonts,amsmath,amssymb]{revtex4}

\usepackage{graphicx}% Include figure files
\usepackage{bm}% bold math
\usepackage{color}
\usepackage{multirow}
\usepackage{slashed}

\begin{document}
\baselineskip=17pt \parskip=5pt

\preprint{NCTS-PH/1809}
\hspace*{\fill}

\title{Decay rate and asymmetries of \,\boldmath$\Sigma^+\to p\mu^+\mu^-$}

\author{Xiao-Gang He$^{1,2,3}$}
\author{Jusak Tandean$^{2,3}$}
\author{German Valencia$^4$}

\affiliation{
$^1$Tsung-Dao Lee Institute $\&$ Department of Physics and Astronomy, SKLPPC, Shanghai Jiao Tong University,
800 Dongchuan Rd., Minhang, Shanghai 200240, China \smallskip \\
$^2$Department of Physics and Center for Theoretical Sciences, National Taiwan
University,\\ No.\,\,1, Sec.\,\,4, Roosevelt Rd., Taipei 106, Taiwan \smallskip \\
$^3$Physics Division, National Center for Theoretical Sciences,
No.\,\,101, Sec.\,\,2, Kuang Fu Rd., Hsinchu 300, Taiwan \smallskip \\
$^4$School of Physics and Astronomy, Monash University, Melbourne VIC-3800, Australia
\bigskip}

%\date{\vspace*{3ex}}

\begin{abstract}

The LHCb Collaboration has recently reported evidence for the rare hyperon decay \,$\Sigma^+\to p\mu^+\mu^-$\, that is consistent with the standard model expectation.
Motivated by this new result we revisit the calculation of this mode including both long and short distance contributions. In the standard model this mode is completely dominated by long distance physics and thus subject to large uncertainties. We present a range of predictions for the rate, the dimuon invariant mass spectrum, and a muon polarization asymmetry that covers these uncertainties as well as known constraints.  We study the interplay between short and long distance contributions which could result in additional asymmetries, but we find that they are negligible within the standard model. We propose a parameterization of these asymmetries in terms of a couple of constants that can arise from new physics.

\end{abstract}

\maketitle

\section{Introduction \label{intro}}

The rare hyperon decay \,$\Sigma^+\to p\mu^+\mu^-$\, gained much attention in 2005 when the HyperCP Collaboration at
Fermilab reported the first evidence for this mode consisting of three events with the same dimuon invariant mass
\,$M_{\mu\mu}=214.3$ MeV \cite{Park:2005ek}. Although the implied rate was consistent with its standard model (SM) value,
the event clustering hinted at the possibility that the decay was mediated by a light new particle.

The LHCb Collaboration, now in a position to repeat that measurement with a much larger event sample, has recently already
turned up $4.1\sigma$ evidence for the decay \cite{Aaij:2017ddf}. The reported branching fraction,
\begin{equation}
{\cal B}\big(\Sigma^+\to p\mu^+\mu^-\big)_{\rm LHCb} \,=\, \big(2.2^{+1.8}_{-1.3}\big)\times10^{-8} \,,
\end{equation}
is in agreement with the range implied by the HyperCP observation \cite{Park:2005ek},
\begin{equation}
{\cal B}\big(\Sigma^+\to p\mu^+\mu^-\big)_{\rm HyperCP} \,=\, \big(8.6^{+6.6}_{-5.4}\pm5.5\big)\times10^{-8} \,,
\end{equation}
and with the SM prediction  \cite{He:2005yn}
\begin{equation} \label{B2005}
1.6\times10^{-8} \,\le\, {\cal B}\big(\Sigma^+\to p\mu^+\mu^-\big)_{\rm SM} \,\le\, 9.0\times10^{-8} \,.
\end{equation}

Unlike HyperCP's finding, the new data from LHCb do not reveal any significant structure in the dimuon mass spectrum \cite{Aaij:2017ddf}.
If this result is confirmed with the larger event sample they expect to collect in the near future~\cite{Santos:2018zbz}, it will of course close the window on the possibility of a new light particle which was much speculated following the HyperCP announcement~\cite{He:2005we,hypercpx,exp}.
A larger sample of events would at the same time permit a more detailed study of the SM predictions.

With this in mind, we revisit our SM calculation to present up-to-date predictions for the rate and the $M_{\mu\mu}$ distribution.
Moreover, we propose a series of asymmetries related to the muons in this mode to probe it further.
As some of the asymmetries turn out to be tiny in the SM, we also explore how new physics (NP) beyond it might enhance them to detectable levels.

The organization of the paper is as follows. In section~\ref{hamiltonian}, we first briefly review the various SM contributions to \,$\Sigma^+\to p\mu^+\mu^-$\, and subsequently write down the decay amplitude in a form that accommodates them and includes several terms which could be induced by heavy NP.
In section~\ref{observables}, we derive the observables of interest: the rate, the $M_{\mu\mu}$ spectrum, a muon forward-backward asymmetry, and three different polarization asymmetries of the muons.
In section~\ref{smpredict}, we evaluate the SM expectations of these quantities.
Some of them are found to be highly suppressed and therefore may serve as places to look for NP signals.
In section~\ref{bsm}, we consider the potential NP effects which can substantially increase the observables predicted to be tiny in the SM.
In section~\ref{s2pee}, we provide some remarks on the related transition \,$\Sigma^+\to pe^+e^-$.\,
Finally, we conclude in section~\ref{concl}. An appendix contains additional formulas.

\section{Effective Hamiltonian and decay amplitude\label{hamiltonian}}

Within the SM, the amplitude for \,$\Sigma^+\to p\mu^+\mu^-$\, receives short distance (SD) and long distance (LD) contributions.
This process may also be influenced by SD effects due to new physics at high energy scales.

\subsection{SM short distance contribution\label{smsd}}

For this decay the SD physics in the SM is described by the effective Hamiltonian~\cite{Buchalla:1995vs}
\begin{equation} \label{Hsd}
{\cal H}_{\rm eff}^{} \,=\, \frac{G_{\rm F}^{}}{\sqrt2}\, \overline{d}\gamma^\kappa(1-\gamma_5^{})
s\,\overline{\mu} \gamma_\kappa^{} \big( \lambda_u^{}z_{7V}^{}-\lambda_t^{}y_{7V}^{}
- \gamma_5^{} \lambda_t^{}y_{7A}^{} \big) \mu
\,+\, {\rm H.c.} \,,
\end{equation}
where $G_{\rm F}$ is the usual Fermi constant, $z_{7V}$ and $y_{7V,7A}$ are the Wilson coefficients, and \,$\lambda_q^{}=V_{qd}^*V_{qs}^{}$\, with $V_{kl}$ denoting the elements of the Cabibbo-Kobayashi-Maskawa (CKM) matrix.
We have dropped from Eq.\,(\ref{Hsd}) the contribution of the SD electromagnetic-penguin diagram, which can be neglected in strangeness-changing \,$|\Delta S|=1$\, transitions~\cite{Buchalla:1995vs}.
To estimate the SD amplitude ${\cal M}_{\rm SM}^{\rm SD}$ induced by ${\cal H}_{\rm eff}$, we employ the baryonic matrix elements~\cite{He:2005yn,He:2005we}\footnote{The kaon-pole part of $\langle p|\overline{d}\gamma^\nu\gamma_5^{}s|\Sigma^+\rangle$ was not included in \cite{He:2005yn}, but the numerical consequence is negligible due to the smallness of the SD contribution.}
\begin{align}
\langle p|\overline{d}\gamma^\kappa s|\Sigma^+\rangle & \,=\, -\bar u_p^{}\gamma^\kappa u_\Sigma^{} \,, ~~~ ~~
\nonumber \\
\langle p|\overline{d}\gamma^\nu\gamma_5^{}s|\Sigma^+\rangle & \,=\, (D-F) \bigg(
\bar u_p^{}\gamma^\nu\gamma_5^{} u_\Sigma^{} + \frac{m_\Sigma^{}+m_p^{}}{q^2-m_K^2}\,
\bar u_p^{}\gamma_5^{}u_\Sigma^{}\, q^\nu \bigg)
\end{align}
derived from the leading-order strong Lagrangian in chiral perturbation theory ($\chi$PT), where $D$ and $F$ are constants in the Lagrangian and \,$q=p_\Sigma^{}-p_p^{}$\, is the difference between the four-momenta of $\Sigma^+$ and $p$.
Thus,
\begin{align}
{\cal M}_{\rm SM}^{\rm SD} & \,=\, \frac{G_{\rm F}^{}}{\sqrt{2}} \biggl\{ -\bar u_p^{} \gamma^\nu
\big[1+\gamma_5^{}(D-F)\big] u_\Sigma^{}\, \bar u_\mu^{} \gamma_\nu^{} \big(
\lambda_u^{} z_{7V}^{}-\lambda_t^{~} y_{7V}^{}
- \gamma_5^{} \lambda_t^{~} y_{7A}^{} \big) v_{\bar\mu}^{}
\nonumber \\ & \hspace{8ex} +\,
2(D-F) \frac{m_\Sigma^{}+m_p^{}}{q^2-m_K^2}\, \lambda_t^{~} y_{7A\,}^{} m_\mu^{}\,
\bar u_p^{}\gamma_5^{}u_\Sigma^{}\,\bar u_\mu^{}\gamma_5^{}v_{\bar\mu}^{} \biggr\} \,.
\end{align}

\subsection{SM long distance contribution\label{smld}}

The LD contribution arises mainly from the photon-mediated process \,$\Sigma^+\to p\gamma^*\to p\mu^+\mu^-$.\,
The amplitude ${\cal M}_{\rm SM}^{\rm LD}$ for this transition can be parameterized by four (complex) gauge-invariant form factors \cite{Bergstrom:1987wr}.
In the notation of Ref.\,\cite{He:2005yn}
\begin{equation}
{\cal M}_{\rm SM}^{\rm LD} \,=\,
\frac{-i e^2 G_{\rm F}^{}}{q^2}\, \bar u_p^{} (a+b\gamma_5^{})\sigma_{\kappa\nu}^{}
q^\kappa u_\Sigma^{}\, \bar u_\mu^{}\gamma^\nu v_{\bar\mu}^{}
- e^2 G_{\rm F}^{}\, \bar u_p^{}\gamma_\kappa^{}(c+d\gamma_{5}^{}) u_\Sigma^{}\,
\bar u_\mu^{}\gamma^\kappa v_{\bar\mu}^{} \,,
\end{equation}
where \,$e^2=4\pi\alpha(0)$\, with \,$\alpha(0)\simeq1/137$ \cite{pdg} and $a,b,c$, and $d$ represent the form factors, which are functions of \,$q^2=M_{\mu\mu}^2$.\,
Their expressions are available from Ref.\,\cite{He:2005yn}.

\subsection{Amplitude including new physics contributions}

Beyond the SM, new interactions at SD may generate contributions to \,$\Sigma^+\to p\mu^+\mu^-$\, which modify the SM portion of the decay amplitude, $\cal M$, and/or give rise to new terms in it.
To accommodate both cases, in this paper we consider the form
\begin{align} \label{MtotS2pll}
{\cal M} \,= &~ \bigl[ i q_\kappa^{}\, \bar u_p^{} \big( \tilde{\textsc a}
+ \gamma_5^{} \tilde{\textsc b} \big) \sigma^{\nu\kappa} u_\Sigma^{}
- \bar u_p^{}\gamma^\nu \big(\tilde{\textsc c}+\gamma_5^{}\tilde{\textsc d}
\big) u_\Sigma^{} \bigr] \bar u_\mu^{}\gamma_\nu^{} v_{\bar\mu}^{}
\,+\, \bar u_p^{}\gamma^\nu \big(\tilde{\textsc e}+\gamma_5^{}\tilde{\textsc f}
\big) u_\Sigma^{}\, \bar u_\mu^{} \gamma_\nu^{} \gamma_5^{} v_{\bar\mu}^{}
\nonumber \\ & +\,
\bar u_p^{}\big(\tilde{\textsc g}+\gamma_5^{}\tilde{\textsc h}\big)u_\Sigma^{}\,
\bar u_\mu^{}v_{\bar\mu}^{}
\,+\, \bar u_p^{}\big(\tilde{\textsc j}+\gamma_5^{}\tilde{\textsc k}\big)u_\Sigma^{}\,
\bar u_\mu^{}\gamma_5^{}v_{\bar\mu}^{} \,,
\end{align}
where \,$q=p_\Sigma^{}-p_p^{}=p_+^{}+p_-^{}$,\, with $p_\pm^{}$ denoting the four-momenta of $\mu^\pm$, and \,$\tilde{\textsc a},\tilde{\textsc b},.$..$,\tilde{\textsc k}$\, are complex coefficients.
As some of the observables to be treated in the next section are predicted to be very small in the SM, later on we will entertain the possibility that NP affects some of the coefficients and brings about substantially enhancing effects on such observables.

\section{Observables\label{observables}}

After averaging (summing) $|{\cal M}|^2$ over initial (final) spins, we arrive at the double-differential decay rate
\begin{align} \label{d^2G/ds/dt}
\frac{d^2\Gamma(\Sigma^+\to p\mu^+\mu^-)}{d q^2\, dt} \,=\,
\frac{\overline{|{\cal M}|^2}}{4(4\pi m_\Sigma)^3} \,,
\end{align}
where \,$t=\big(p_\Sigma^{}-p_-^{}\big)^2=\big(p_+^{}+p_p\big)^2$\, and
\begin{align} \label{|M|^2}
\overline{|{\cal M}|^2} \,=\, &~ 4 \Big[ \big(f_-^{}{\texttt M}_+^2
+ 2 f q^2\big) |\tilde{\textsc a}|^2 +
\big(f_+^{}{\texttt M}_-^2+2f q^2\big) |\tilde{\textsc b}|^2
+ (f_-^{}-2 f)\, |\tilde{\textsc c}|^2 + (f_+^{}-2 f)\, |\tilde{\textsc d}|^2 \Big]
\nonumber \\ & +\,
4 \Big[ \big( \beta^2 \hat m_-^2 q^2+2 \hat m_+^2 m_\mu^2-2 f \big) |\tilde{\textsc e}|^2
+ \big( \beta^2 \hat m_+^2 q^2+2 \hat m_-^2 m_\mu^2-2 f \big) |\tilde{\textsc f}|^2 \Big]
\nonumber \\ & +\,
2 \Big( \beta^2 |\tilde{\textsc g}|^2 \hat m_+^2 + \beta^2 |\tilde{\textsc h}|^2 \hat m_-^2
+ |\tilde{\textsc j}|^2 \hat m_+^2 + |\tilde{\textsc k}|^2 \hat m_-^2 \Big) q^2
\nonumber \\ & +\,
8\, {\rm Re} \Big[ \tilde{\textsc a}{}^*\tilde{\textsc c}\, f_-^{} {\texttt M}_+^{}
- \tilde{\textsc b}{}^*\tilde{\textsc d}\, f_+^{} {\texttt M}_-^{}
+ \big( \tilde{\textsc e}{}^*\tilde{\textsc j}\, \hat m_+^2 {\texttt M}_-^{}
- \tilde{\textsc f}{}^*\tilde{\textsc k}\, \hat m_-^2 {\texttt M}_+^{} \big) m_\mu^{} \Big]
\nonumber \\ & +\,
4 \big(4 m_\mu^2 + \hat m_-^2 + \hat m_+^2 - 4 t\big)\, {\rm Re} \Big[ \big(
\tilde{\textsc a}{}^*\tilde{\textsc g}\, q^2 + \tilde{\textsc b}{}^*\tilde{\textsc h}\, q^2
+ \tilde{\textsc c}{}^* \tilde{\textsc g}\, {\texttt M}_+^{}
- \tilde{\textsc d}{}^* \tilde{\textsc h}\, {\texttt M}_-^{} \big) m_\mu^{}
\nonumber \\ & \hspace{56mm}
- \big( \tilde{\textsc a}{}^* \tilde{\textsc f}\, {\texttt M}_+^{}
- \tilde{\textsc b}{}^* \tilde{\textsc e}\, {\texttt M}_-^{}
+ \tilde{\textsc c}{}^*\tilde{\textsc f}+\tilde{\textsc d}{}^*\tilde{\textsc e} \big) q^2 \Big] \,,
\end{align}
with
\begin{align} \label{ft}
f_\pm^{} & \,=\, \big(2m_\mu^2+q^2\big)\hat m_\pm^2 \,, ~~~~~
\hat m_\pm^2 \,=\, {\texttt M}_\pm^2-q^2 \,, ~~~~~
{\texttt M}_\pm^{} \,=\, m_\Sigma^{}\pm m_p^{} \,, ~~~~ ~~~
\beta \,=\, \sqrt{1-\frac{4m_\mu^2}{q^2}} \,, ~~~~~
\nonumber \\
f & \,=\, \big(m_\Sigma^2+m_p^2+m_\mu^2-q^2-t\big) \big(m_\mu^2-t\big)+m_\Sigma^2 m_p^2 \,.
\end{align}
The allowed range of $t$ is given by
\begin{align}
t_{\rm max, min}^{} \,=\, \mbox{$\frac{1}{2}$} \Big( m_\Sigma^2+m_p^2+2 m_\mu^2-q^2
\pm \beta \sqrt{\bar\lambda} \Big) \,, ~~~~ ~~~
\bar\lambda \,=\, \hat m_-^2 \hat m_+^2 \,.
\end{align}
For \,$\tilde{\textsc a},\tilde{\textsc b},.$..$,\tilde{\textsc k}$\, being independent of $t$, we can integrate Eq.\,(\ref{d^2G/ds/dt}) over $t$ to obtain
\begin{align} \label{dG/ds}
\frac{d\Gamma(\Sigma^+\to p\mu^+\mu^-)}{dq^2} & =
\frac{\big(3\beta-\beta^3\big)q^2\sqrt{\bar\lambda}}{64 \pi^3 m^3_\Sigma} \Bigg\{ \Bigg[
\frac{\hat m_+^2}{3}+\frac{q^2}{2} \Bigg] |\tilde{\textsc a}|^2
+ \Bigg[ 1+\frac{\hat m_+^2}{3q^2} \Bigg] \frac{|\tilde{\textsc c}|^2}{2}
+ {\texttt M}_+^{}\, {\rm Re}(\tilde{\textsc a}{}^*\tilde{\textsc c}) \Bigg\} \hat m_-^2
\nonumber \\ &
+ \frac{\big(3\beta-\beta^3\big)q^2\sqrt{\bar\lambda}}{64 \pi^3 m^3_\Sigma} \Bigg\{ \Bigg[
\frac{\hat m_-^2}{3}+\frac{q^2}{2} \Bigg] |\tilde{\textsc b}|^2
+ \Bigg[ 1+\frac{\hat m_-^2}{3q^2} \Bigg] \frac{|\tilde{\textsc d}|^2}{2}
- {\texttt M}_-^{}\, {\rm Re}(\tilde{\textsc b}{}^*\tilde{\textsc d}) \Bigg\} \hat m_+^2
\nonumber \\ &
+ \frac{\beta\, \sqrt{\bar\lambda}}{32 \pi^3 m_\Sigma^3} \Bigg\{ \Bigg[
\frac{3-\beta^2}{12} \bar\lambda + \frac{\beta^2}{2} \hat m_-^2q^2 + \hat m_+^2 m_\mu^2 \Bigg]
|\tilde{\textsc e}|^2
+ \Big[ \beta^2 |\tilde{\textsc g}|^2+|\tilde{\textsc j}|^2 \Big] \frac{\hat m_+^2 q^2}{4}
\nonumber \\ & \hspace{5em}
+ \Bigg[ \frac{3-\beta^2}{12} \bar\lambda + \frac{\beta^2}{2} \hat m_+^2 q^2
+ \hat m_-^2 m_\mu^2 \Bigg] |\tilde{\textsc f}|^2
+ \Big[ \beta^2 |\tilde{\textsc h}|^2+|\tilde{\textsc k}|^2 \Big] \frac{\hat m_-^2 q^2}{4}
\nonumber \\ & \hspace{5em}
+ \hat m_+^2  m_\mu^{}\, {\texttt M}_-^{}\, {\rm Re}(\tilde{\textsc e}{}^* \tilde{\textsc j})
- \hat m_-^2  m_\mu^{}\, {\texttt M}_+^{}\, {\rm Re}(\tilde{\textsc f}{}^* \tilde{\textsc k})
\Bigg\} \,.
\end{align}

The quantities pertinent to \,$\Sigma^+\to p\mu^+\mu^-$\, that have been measured so far are the branching fraction and the dimuon invariant-mass distribution~\cite{Park:2005ek,Aaij:2017ddf}.
If a good amount of data on this decay become available from future experimental efforts, there are other observables that can be analyzed.

One of them is the forward-backward asymmetry ${\cal A}_{\rm FB}$ of the muon defined as
\begin{equation}
{\cal A}_{\rm FB}^{} \,=\, \frac{\int_{-1}^1dc_\theta^{}~{\rm sgn}(c_\theta^{})~ \Gamma''}
{\int_{-1}^1dc_\theta^{}~\Gamma''} \,, ~~~ ~~~~
\Gamma'' \,\equiv\, \frac{d^2\Gamma(\Sigma^+\to p\mu^+\mu^-)}{dq^2\, dc_\theta^{}} \,, ~~~ ~~
c_\theta^{} \,\equiv\, \cos\theta \,,
\end{equation}
where $\theta$ is the angle between the directions of $\mu^-$ and $p$ in the rest frame
of the dimuon system.
As outlined in appendix \ref{app}, from Eqs.\,\,(\ref{d^2G/ds/dt}) and (\ref{|M|^2}) we can get
\begin{equation}
\frac{d^2\Gamma(\Sigma^+\to p\mu^+\mu^-)}{dq^2\, dc_\theta^{}} \,=\,
{\cal F}_0^{} + {\cal F}_1^{}\, c_\theta^{} + {\cal F}_2^{}\, c_\theta^2 \,,
\end{equation}
with ${\cal F}_{0,1,2}$ being independent of $\theta$ and written down in Eq.\,(\ref{F2}),
leading to
\begin{equation} \label{AFB}
{\cal A}_{\rm FB}^{} \,=\, \frac{\beta^2\, \bar\lambda}{64\pi^{3\,}\Gamma'\, m_\Sigma^3}\,
{\rm Re}\! \begin{array}[t]{l} \Big\{ \big[ {\texttt M}_+^{} \tilde{\textsc a}{}^*\tilde{\textsc f}
- {\texttt M}_-^{} \tilde{\textsc b}{}^* \tilde{\textsc e}
- \big( \tilde{\textsc a}{}^*\tilde{\textsc g}+\tilde{\textsc b}{}^*\tilde{\textsc h}\big) m_\mu
+ \tilde{\textsc c}{}^*\tilde{\textsc f} + {\textsc d}{}^*\tilde{\textsc e} \big] q^2
\vspace{1pt} \\ \;\,
-\; \big( {\texttt M}_+^{} \tilde{\textsc c}{}^* \tilde{\textsc g}
- {\texttt M}_-^{} \tilde{\textsc d}{}^* \tilde{\textsc h} \big) m_\mu \Big\} \,, \end{array}
\end{equation}
with
\begin{equation}
\Gamma' \,\equiv\, \frac{d\Gamma(\Sigma^+\to p\mu^+\mu^-)}{dq^2} \,=\,
\int_{-1}^1dc_\theta^{}~\Gamma'' \,.
\end{equation}
This may be the main observable that could provide a window into NP interactions modifying the terms in the SM amplitude not dominated by the LD components.
Also of interest is the integrated forward-backward asymmetry
\begin{equation}
\tilde A_{\rm FB}^{} \,=\,\frac{1}{\Gamma(\Sigma^+\to p\mu^+\mu^-)}
\int_{q_{\rm min}^2}^{q_{\rm max}^2}dq^2\int_{-1}^1dc_\theta^{}~{\rm sgn}(c_\theta^{})~\Gamma'' \,,
\end{equation}
where
\begin{equation}
\Gamma(\Sigma^+\to p\mu^+\mu^-) \,=\, \int_{q_{\rm min}^2}^{q_{\rm max}^2}dq^2\,\Gamma' \,, ~~~ ~~~~
q_{\rm min}^2 \,=\, 4m^2_\mu \,, ~~~~~ q_{\rm max}^2 \,=\, \big(m_\Sigma^{}-m_p\big)^2 \,.
\end{equation}

In the rest frame of the dimuon system, we can also investigate the polarization asymmetries of
the leptons.\footnote{The forward-backward and polarization asymmetries which we consider are
analogous to those previously addressed in the literature on the inclusive decay
\,$b\to s\ell^+\ell^-$ \cite{b2sll} and the baryon decay
\,$\Lambda_b\to\Lambda\ell^+\ell^-$ \cite{Lb2Lll}, as well as the kaon decay
\,$K^+\to\pi^+\ell^+\ell^-$ \cite{Savage:1990km}.}
In this frame, the four-momenta of the particles involved and $\theta$ are given by
\begin{align} \label{pc}
p_\Sigma^{} & \,=\, \big(E_\Sigma^{},\bm{p}_\Sigma^{}\big) \,, ~~~~~ ~~~~~
p_-^{} \,=\, \big(E_\mu,\bm{p}_\mu\big) \,, & p_+^{} & \,=\, \big(E_\mu,-\bm{p}_\mu\big) \,, &
\nonumber \\
p_p^{} & \,=\, \big(E_p,\bm{p}_p\big) \,=\, \big(E_\Sigma^{}-2E_\mu,\bm{p}_\Sigma^{}\big) \,, &
c_\theta^{} & \,=\, \frac{\bm{p}_\mu^{}\cdot\bm{p}_p^{}}
{\big|\bm{p}_\mu\big|\,\big|\bm{p}_p\big|} \,. &
\end{align}
Furthermore, we can select the unit vectors of the coordinate axes to be
\begin{equation}
\hat{\bm{z}} \,=\, \frac{\bm{p}_\mu^{}}{\big|\bm{p}_\mu\big|} \,, ~~~~ ~~~~
\hat{\bm{y}} \,=\,
\frac{\bm{p}_p^{}\times\bm{p}_\mu^{}}{\big|\bm{p}_p\times\bm{p}_\mu\big|} \,, ~~~~ ~~~~
\hat{\bm{x}} \,=\, \hat{\bm{y}}\!\times\!\hat{\bm{z}} \,,
\end{equation}
and so the particles move only on the $x$-$z$ plane.
Accordingly, the spin four-vectors of $\mu^\pm$ are
\begin{align}
s_\pm^{} \,=\, \bigg(\frac{\mp|\bm{p}_\mu|_{\,} \varsigma_z^\pm}{m_\mu},\,
\varsigma_x^\pm\hat{\bm{x}} + \varsigma_y^\pm\hat{\bm{y}} +
\frac{E_\mu}{m_\mu}\, \varsigma_z^\pm\hat{\bm{z}} \bigg) \,,
\end{align}
where $\varsigma_{x,y,z}^\pm$ are the spin three-vector components satisfying
\,$(\varsigma_x^\pm)^2+(\varsigma_y^\pm)^2+(\varsigma_z^\pm)^2=1$.\,
Thus, $s_\pm^{}$ fulfill the normalization and orthogonality requirements \,$s_\pm^2=-1$\, and
\,$s_\pm^{}\cdot p_\pm^{}=0$\, and reduce to \,$s_\pm^{}=\big(0,\varsigma_x^\pm\hat{\bm{x}}
+\varsigma_y^\pm\hat{\bm{y}}+\varsigma_z^\pm\hat{\bm{z}}\big)$\,
in the rest frames of $\mu^\pm$, respectively.
These choices lead to
\begin{align} \label{p.s}
p_\Sigma^{}\cdot s_\mp^{} & \,=\, \frac{\pm\beta_{\,}\big(m_\Sigma^2-m_p^2+q^2\big)
- \sqrt{\bar\lambda}\, c_\theta^{}}{4 m_\mu^{}}\, \varsigma_z^\mp
+ \sqrt{\frac{\bar\lambda}{q^2}}~ \frac{s_{\theta\,}^{}\varsigma_x^\mp}{2} \,,
\nonumber \\
p_\pm^{}\cdot s_\mp^{} & \,=\, \pm \frac{\beta q^2\, \varsigma_z^\mp}{2 m_\mu^{}} \,,
~~~~ ~~~~ \epsilon_{\eta\kappa\nu\rho}^{}\,p_{\Sigma\,}^\eta p_-^\kappa p_+^\nu s_\mp^\rho
\,=\, \frac{\beta}{4}\sqrt{\bar\lambda_{\,}q^2}~ s_\theta^{}\, \varsigma_y^\mp \,,
\end{align}
with \,$s_\theta^{}=\sqrt{1-c_\theta^2}$\, and \,$\epsilon_{0123}^{}=+1$.\,
To deal with a specific polarization of $\mu^-$ $(\mu^+)$, we change $u_\mu$ $(v_{\bar\mu})$ in
${\cal M}$ to \,$\hat\Sigma_-u_\mu$ $\big(\hat\Sigma_+v_{\bar\mu}\big)$
with the spin projection operator
\,$\hat\Sigma_{-(+)}=\tfrac{1}{2} \big( 1+\gamma_5^{}\slashed s{}_{-(+)} \big)$\,
and evaluate $|{\cal M}|^2$ with the aid of Eq.\,(\ref{p.s}).

To examine the decay when $\mu^-$ is polarized, we can express the resulting differential rate as
\begin{align} \label{dGxyz/ds}
\frac{d\Gamma^-(\varsigma_x^-,\varsigma_y^-,\varsigma_z^-)}{dq^2} \,=\,
\frac{\Gamma'}{2} \big( 1 \,+\, {\cal P}_{\rm T\,}^- \varsigma_x^-
+ {\cal P}_{\rm N\,}^- \varsigma_y^- + {\cal P}_{\rm L\,}^- \varsigma_z^- \big) \,,
\end{align}
where the subscripts L, N, and T refer to the longitudinal, normal, and transverse polarizations
of $\mu^-$ along the chosen $z$-, $y$-, and $x$-axes, respectively.
We find the corresponding polarization asymmetries ${\cal P}_{\rm L,N,T}^-$ to be
\begin{align} \label{PLNT-}
{\cal P}_{\rm L}^- \,=\;\, & \frac{\beta^2\, \sqrt{\bar\lambda}}{192\pi^{3\,}\Gamma'\,m_\Sigma^3}\,
{\rm Re}\! \begin{array}[t]{l} \Big\{ \bigl[ -3 \big(
2 {\texttt M}_+^{} \tilde{\textsc a}{}^* \tilde{\textsc e}
+ \tilde{\textsc h}{}^* \tilde{\textsc k} \big) q^2
- 2 \big(\hat m_+^2+3q^2\big) \tilde{\textsc c}{}^*\tilde{\textsc e}
+ 6 m_\mu {\texttt M}_+^{} \tilde{\textsc f}{}^*\tilde{\textsc h} \bigr] \hat m_-^2
\vspace{2pt} \\ \;\,
+\; \big[ 3 \big( 2 {\texttt M}_-^{} \tilde{\textsc b}{}^*\tilde{\textsc f}
- \tilde{\textsc g}{}^* \tilde{\textsc j} \big) q^2
- 2 \big(\hat m_-^2+3q^2\big) \tilde{\textsc d}{}^*\tilde{\textsc f}
- 6 m_\mu {\texttt M}_-^{} \tilde{\textsc e}{}^*\tilde{\textsc g}
\big] \hat m_+^2 \Big\} \,, \end{array} \vphantom{\int_{\int_|^|}^{}}
\nonumber \\
{\cal P}_{\rm N}^- \,=\;\, & \frac{\beta^2\, \bar\lambda\, \sqrt{q^2}}{256\pi^{2\,} \Gamma'\,
m_\Sigma^3}\, {\rm Im}\! \begin{array}[t]{l} \Big\{ 2 \big[
( {\texttt M}_+^{} \tilde{\textsc a}+\tilde{\textsc c}{} )^* \tilde{\textsc f}
+ ( \tilde{\textsc d}-{\texttt M}_-^{}\tilde{\textsc b} )^* \tilde{\textsc e} \big] m_\mu
- \big(\tilde{\textsc a}{}^*\tilde{\textsc g} + \tilde{\textsc b}{}^*\tilde{\textsc h}\big) q^2
\vspace{2pt} \\ \;\,
-\; \big(\tilde{\textsc c}{}^*\tilde{\textsc g}-\tilde{\textsc e}{}^*\tilde{\textsc j}\big)
{\texttt M}_+^{} + \big(\tilde{\textsc d}{}^*\tilde{\textsc h} - \tilde{\textsc f}{}^*
\tilde{\textsc k}\big) {\texttt M}_-^{} \Big\} \,, \end{array} \vphantom{\int_{\int_|^|}^{}}
\nonumber \\
{\cal P}_{\rm T}^- \,=\;\, & \frac{\beta\, \bar\lambda\, \sqrt{q^2}}{256\pi^{2\,} \Gamma'\,
m_\Sigma^3}\, {\rm Re}\! \begin{array}[t]{l} \Big\{ 2 \big[ 2 \big( {\texttt M}_+^{}
\tilde{\textsc a} + \tilde{\textsc c}{} \big)^{\!*} \big( \tilde{\textsc d}-{\texttt M}_-^{}
\tilde{\textsc b} \big) - {\texttt M}_-^{} \tilde{\textsc a}{}^*\tilde{\textsc e}
+ {\texttt M}_+^{} \tilde{\textsc b}{}^* \tilde{\textsc f} \big] m_\mu
\vspace{2pt} \\ \;\,
-\;\, {\texttt M}_+^{} \tilde{\textsc c}{}^* \tilde{\textsc j}
+ {\texttt M}_-^{} \tilde{\textsc d}{}^* \tilde{\textsc k}
+ \beta^2 \big( {\texttt M}_+^{} \tilde{\textsc e}{}^* \tilde{\textsc g}
- {\texttt M}_-^{} \tilde{\textsc f}{}^* \tilde{\textsc h} \big) \Big\} \end{array}
\nonumber \\ &
-\, \frac{\beta\, \bar\lambda~ {\rm Re} \Big[ \big( \tilde{\textsc a}{}^* \tilde{\textsc j}
+ \tilde{\textsc b}{}^*\tilde{\textsc k} \big) q^4 + 2 \big( \tilde{\textsc c}{}^*\tilde{\textsc e}
+ \tilde{\textsc d}{}^* \tilde{\textsc f} \big) {\texttt M}_+^{} {\texttt M}_-^{} m_\mu
\Big]}{256\pi^{2\,}\Gamma'\, m_\Sigma^3 \sqrt{q^2}} \,.
\end{align}
For $\mu^+$ being polarized, we can obtain formulas analogous to Eqs.\,\,(\ref{dGxyz/ds})-(\ref{PLNT-}).
With respect to their $\mu^-$ counterparts, the resulting polarization asymmetries are
\begin{align} \label{PLNT+}
{\cal P}_{\rm L}^+ & \,=\, {\cal P}_{\rm L}^- + \frac{\beta^2 \sqrt{\bar\lambda}~ {\rm Re} \Big[
2 \big( \hat m_+^2 {\texttt M}_-^{} \tilde{\textsc e}{}^* \tilde{\textsc g}
- \hat m_-^2 {\texttt M}_+^{} \tilde{\textsc f}{}^* \tilde{\textsc h} \big) m_\mu
+ \big( \hat m_+^2 \tilde{\textsc g}{}^* \tilde{\textsc j} + \hat m_-^2 \tilde{\textsc h}{}^*
\tilde{\textsc k} \big) q^2 \Big]}{32\pi^{3\,} \Gamma'\, m_\Sigma^3} \,,
\nonumber \\
{\cal P}_{\rm N}^+ & \,=\, {\cal P}_{\rm N}^- + \frac{\beta^2 \bar\lambda~ {\rm Im} \big(
{\texttt M}_-^{} \tilde{\textsc f}{}^* \tilde{\textsc k} - {\texttt M}_+^{} \tilde{\textsc e}{}^*
\tilde{\textsc j} \big) \sqrt{q^2}}{128\pi^{2\,} \Gamma'\, m_\Sigma^3} \,,
\nonumber \\
{\cal P}_{\rm T}^+ & \,=\, {\cal P}_{\rm T}^-
+ \frac{\beta \bar\lambda~ {\rm Re} \Big[ \big( \tilde{\textsc a} q^2
+ {\texttt M}_+^{} \tilde{\textsc c} \big)^{\!*} \big( 2 m_\mu {\texttt M}_-^{} \tilde{\textsc e}
+ \tilde{\textsc j} q^2\big)
- \big( \tilde{\textsc b} q^2 - {\texttt M}_-^{} \tilde{\textsc d} \big)^{\!*} \big( 2 m_\mu
{\texttt M}_+^{} \tilde{\textsc f} - \tilde{\textsc k} q^2 \big) \Big]}
{128\pi^{2\,} \Gamma'\, m_\Sigma^3 \sqrt{q^2}} \,.
\end{align}
Similarly to the forward-backward asymmetry, we can also define the integrated $\mu^\mp$
polarization asymmetries
\begin{equation}
\tilde P_{\rm L,N,T}^\mp \,=\, \frac{1}{\Gamma(\Sigma^+\to p\mu^+\mu^-)}
\int_{q_{\rm min}^2}^{q_{\rm max}^2}dq^2~ \Gamma'\, {\cal P}_{\rm L,N,T}^\mp \,.
\end{equation}

We notice that ${\cal P}_{\rm L}^\pm$ probe parity violation in the leptonic parts of the amplitude, while ${\cal P}_{\rm T}^\pm$ are sensitive to parity violation in the hadronic and/or leptonic parts of the amplitude.
On the other hand, ${\cal P}_{\rm N}^\pm$  are (naive-)$T$-odd, being proportional to
\,$\epsilon_{\eta\kappa\nu\rho}^{}\, p_{\Sigma\,}^\eta p_-^\kappa p_+^\nu s_\pm^\rho$,\, and can be induced by $CP$-violating new physics or by unitarity phases which in this case are large.

\section{Standard model predictions\label{smpredict}}

The nonzero coefficients in Eq.\,(\ref{MtotS2pll}) due to the SD and LD amplitudes in the SM alone are
\begin{align} \label{abcdefk}
\tilde{\textsc a} & \,=\, \frac{e^2 G_{\rm F}^{}\,a}{q^2} \,, &
\tilde{\textsc b} & \,=\, \frac{e^2 G_{\rm F}^{}\,b}{q^2} \,,
\nonumber \\
\tilde{\textsc c} & \,=\, e^2 G_{\rm F}^{}\, c
+ G_{\rm F}^{}\, \frac{\lambda_u^{} z_{7V}^{}-\lambda_t^{~} y_{7V}^{}}{\sqrt2} \,, &
\tilde{\textsc d} & \,=\, e^2 G_{\rm F}^{}\, d
+ \frac{D-F}{\sqrt2}\,G_{\rm F}^{} \big(\lambda_u^{} z_{7V}^{}-\lambda_t^{~} y_{7V}^{}\big) \,,
\nonumber \\
\tilde{\textsc e} & \,=\, \frac{G_{\rm F}^{}}{\sqrt2}\, \lambda_t^{~} y_{7A}^{} \,, &
\tilde{\textsc f} & \,=\, \frac{D-F}{\sqrt2}\, G_{\rm F\,}^{} \lambda_t^{~} y_{7A}^{} \,,
\nonumber \\
\tilde{\textsc k} & \,=\, \frac{m_\Sigma^{}+m_p^{}}{q^2-m_K^2}\sqrt2\, (D-F) G_{\rm F\,}^{}
\lambda_t^{~}y_{7A}^{}m_\mu^{}  & &
\end{align}
from subsections \ref{smsd} and \ref{smld}.
To assess their effects on our observables of concern, we adopt \,$z_{7V}^{}=-0.046\alpha$,\, $y_{7V}^{}=0.73\alpha$,\, and
\,$y_{7A}^{}=-0.654\alpha$~\cite{Buchalla:1995vs,Mescia:2006jd}, with \,$\alpha=\alpha(m_Z)\simeq1/129$~\cite{Davier:2017zfy}, the up-to-date CKM parameters supplied by Ref.\,\cite{pdg}, and \,$D=0.81$\, and \,$F=0.46$\, from fitting the lowest-order matrix elements of charged currents in semileptonic baryon decays~\cite{Bijnens:1985kj} to their existing data~\cite{pdg}.

In Ref.\,\cite{He:2005yn}, to determine the $q^2$ dependences of the form factors $a,b,c$, and $d$, first we calculated their imaginary parts with the aid of unitarity arguments.
Subsequently, with the resulting Im$(a,b)$ evaluated at $q^2=0$, we approximated the real parts of $a$ and $b$ as constants which were fixed from the measured rate and asymmetry parameter of the radiative weak decay \,$\Sigma^{+}\to p\gamma$,\, up to a fourfold ambiguity.
Moreover, since Im$(a,b,c,d)$ could be treated in either the relativistic or heavy-baryon approach of $\chi$PT, we took both of them and as a consequence ended up with eight sets of Re$(a,b)$ values.
In the present paper, employing the updated $D$ and $F$ values given above plus the latest pion-decay constant \,$f_\pi^{}=92.07\;$MeV~\cite{pdg},\, we repeat these steps, which lead to \,Im$\big(a(0),b(0)\big)=(2.85,-1.84)$\,MeV\, and \,$(6.86,-0.54)$\,MeV\, in the relativistic and heavy-baryon approaches, respectively, as well as to the Re$(a,b)$ numbers listed in Table\,\,\ref{smresults}.
These results are almost unchanged from their counterparts in Ref.\,\cite{He:2005yn}.
As for the real parts of $c$ and $d$, which cannot yet be extracted from experiment, we simply follow the vector-meson-dominance approximation of Ref.\,\cite{He:2005yn}.

Putting together the SD and LD contributions to \,$\Sigma^+\to p\mu^+\mu^-$,\, we evaluate its branching fraction $\cal B$ and collect the results for the 8 different sets of Re$(a,b)$ in the third column of Table\,\,\ref{smresults}, where for the first [last] 4 rows the relativistic [heavy baryon] expressions for Im$(a,b,c,d)$ have been used.
These $\cal B$ numbers are again little changed from those in Ref.\,\cite{He:2005yn}, the slight differences between them being partly due to the SD component being neglected therein and partly due to the updated input parameters.
The SD contribution alone would produce \,${\cal B}\sim10^{-12}$~\cite{He:2005yn}.
Compared to the 2$\sigma$ upper-limit of the LHCb finding \,${\cal B}=\big(2.2^{+1.8}_{-1.3}\big)\times10^{-8}$ \cite{Aaij:2017ddf}, the $\cal B$ predictions less than \,${\footnotesize\sim\,}6\times10^{-8}$\, in this table appear to be favored, but the statistical confidence of the data is still too low for a definite conclusion.

\begin{table}[t]
\begin{tabular}{|c|c||c|c|c|c|c|}
\hline \hline
~ $\displaystyle\frac{{\rm Re}\,a}{\mbox{\footnotesize MeV}}\vphantom{\int_|^|}$ ~ &
~ $\displaystyle\frac{{\rm Re}\,b}{\mbox{\footnotesize MeV}}$ ~ & ~$10^{8\,}{\cal B}$~ &
~$10^5\tilde A_{\rm FB}^{}$~ & ~$10^5\tilde P_{\rm L}^-$~ & ~$10^6\tilde P_{\rm N}^-$~ &
~$\tilde P_{\rm T}^-\;(\%)$~
\\ \hline \hline
  ~ 13.3~ & \,$-$6.0~ & 1.6 &  ~ 3.7~ & \,$-$7.0~ & $-$0.2 & 59$\vphantom{|_o^|}$ \\
 $-$13.3~ &   ~~ 6.0~ & 3.5 & $-$1.4~ &  \,~ 4.5~ & $-$9.6 & 50$\vphantom{|_o^|}$  \\
  ~~ 6.0~ &  $-$13.3~ & 5.1 &  ~ 0.9~ & \,$-$5.1~ & $-$1.1 & 23$\vphantom{|_o^|}$  \\
\,$-$6.0~ &   ~ 13.3~ & 9.1 & $-$0.3~ &  \,~ 3.3~ & $-$3.1 & 17$\vphantom{|_o^|}$  \\
\hline
  ~ 11.0~ & \,$-$7.4~ & 2.4 &  ~ 2.7~ & \,$-$5.7~ & $-$7.3 & 41$\vphantom{|_o^|}$  \\
 $-$11.0~ &   ~~ 7.4~ & 4.7 & $-$0.7~ &  \,~ 4.1~ & $-$10  & 36$\vphantom{|_o^|}$  \\
  ~~ 7.4~ &  $-$11.0~ & 4.0 &  ~ 1.4~ & \,$-$5.2~ & $-$5.0 & 26$\vphantom{|_o^|}$  \\
\,$-$7.4~ &   ~ 11.0~ & 7.4 & $-$0.3~ &  \,~ 3.6~ & $-$6.0 & 21$\vphantom{|_o^|}$  \\
\hline \hline
\end{tabular}
\caption{Sample values of the branching fraction $\cal B$ of \,$\Sigma^+\to p\mu^+\mu^-$\, and the corresponding integrated asymmetries $\tilde A_{\rm FB}$ and $\tilde P_{\rm L,N,T}^-$ computed within the SM including the SD and LD contributions.
In the evaluation of the $\cal B$, $\tilde A_{\rm FB}$, and $\tilde P_{\rm L,N,T}^-$ entries in the first [last] four rows, the relativistic [heavy baryon] expressions for Im$(a,b,c,d)$ have been used, as explained in the text.\label{smresults}}
\end{table}

\begin{figure}[t]
\includegraphics[width=65mm]{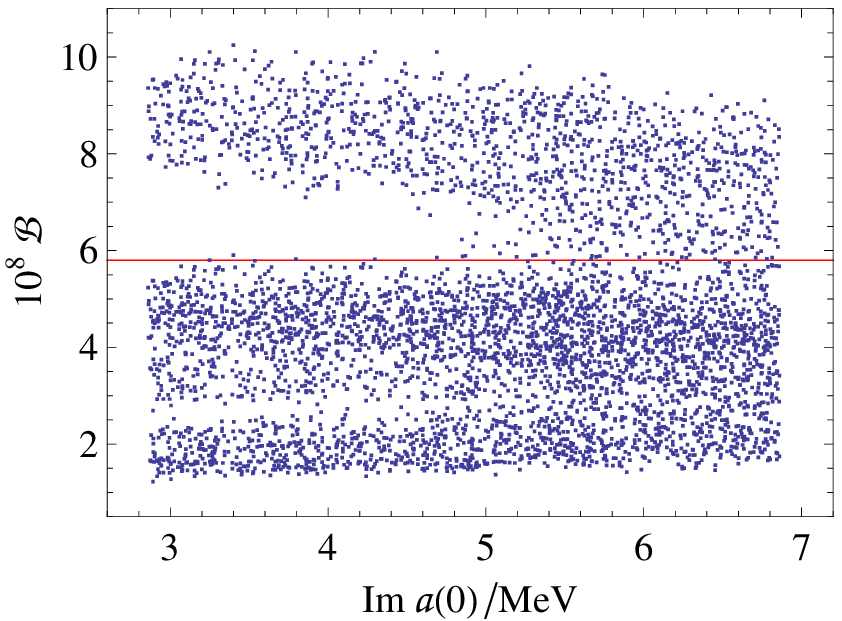} ~
\includegraphics[width=65mm]{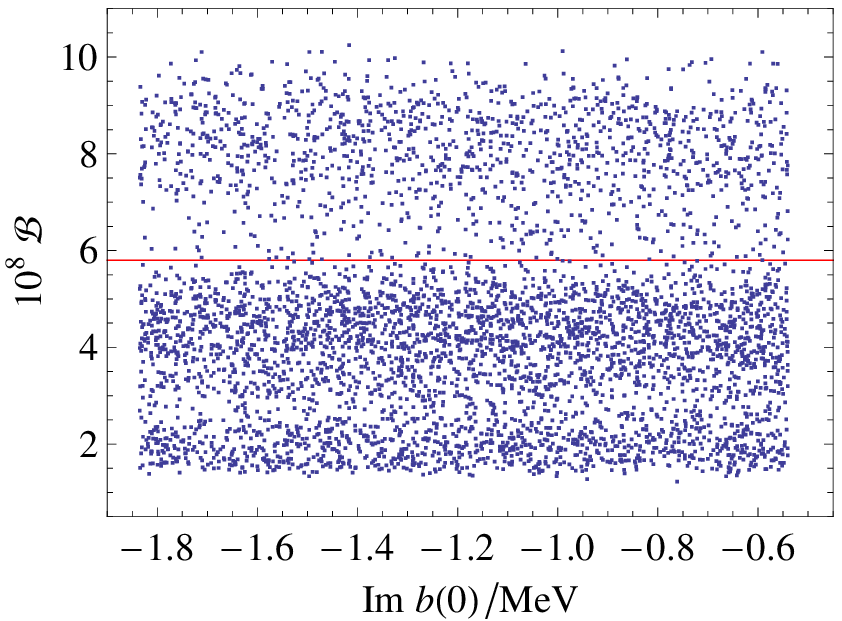} \vspace{7pt} \\
\includegraphics[width=65mm]{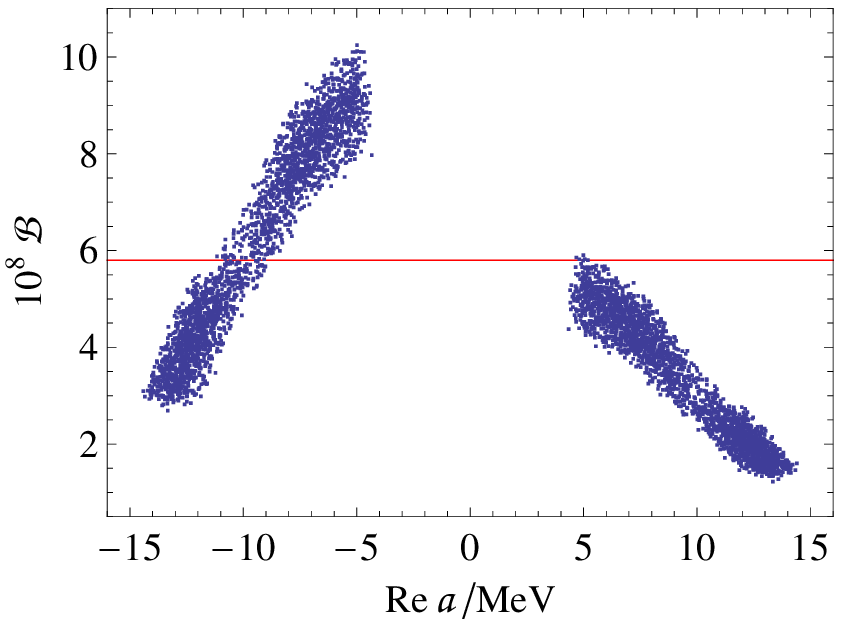} ~
\includegraphics[width=65mm]{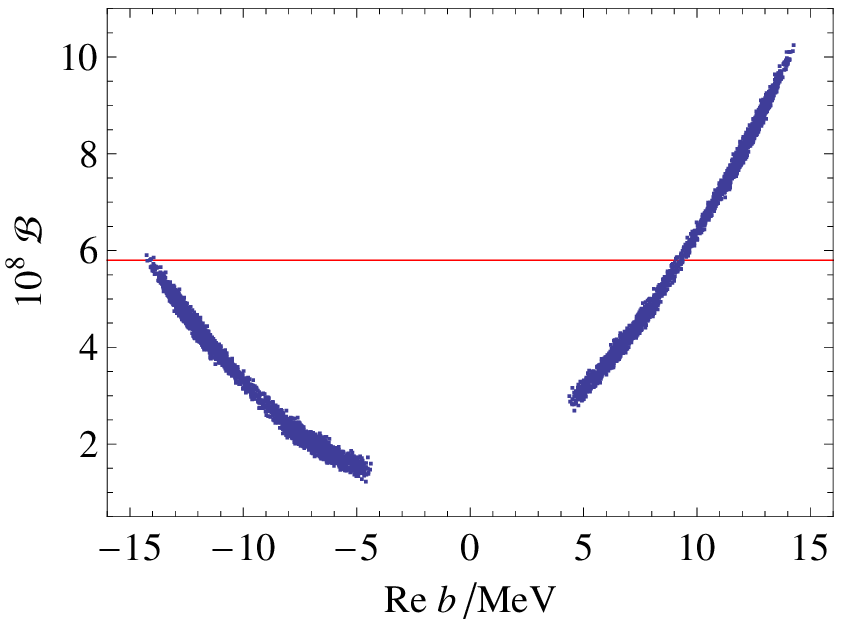} \vspace{-5pt}
\caption{Sample points of \,${\cal B}(\Sigma^+\to p\mu^+\mu^-)\times10^8$\, in relation to the preferred ranges of Im$(a,b)$ at $q^2=0$ and of Re$(a,b)$, as explained in the text.
Each horizontal red line marks the 2$\sigma$ upper-limit of the LHCb measurement~\cite{Aaij:2017ddf}. \label{Bvsab}}
\end{figure}

To explore the predictions further in anticipation of more data from LHCb, in this paper we adopt a slightly different point of view from Ref.\,\cite{He:2005yn}.
We regard the differences between the Im$(a,b,c,d)$ formulas in the relativistic and heavy-baryon approaches as a measure of the theoretical uncertainty in the calculations, attributable to higher orders in $\chi$PT, and present results reflecting the respective ranges of Im$\big(a(0),b(0)\big)$ and Re$(a,b)$.
To do this, we generate randomly several thousand benchmark points satisfying the restrictions from the $\Sigma^+\to p\gamma$ data at the 2$\sigma$ level,\footnote{The pertinent observables are the branching fraction \,${\cal B}(\Sigma^+\to p\gamma)=4\alpha(0)\, G_{\rm F}^2 \big(|a(0)|^2+|b(0)|^2\big) E_\gamma^3 \tau_\Sigma^{}$\, and the decay parameter \,$\alpha_\Sigma^{}=2\,{\rm Re}\big(a^*(0)\,b(0)\big)/\big(|a(0)|^2+|b(0)|^2\big)$,\, where \,$E_\gamma=\big(m_\Sigma^2-m_p^2\big)/(2m_\Sigma^{})$\, and $\tau_\Sigma^{}$ is the $\Sigma^+$ lifetime.
Their experimental values are \,${\cal B}(\Sigma^+\to p\gamma)=(1.23\pm 0.05)\times10^{-3}$\, and \,$\alpha_\Sigma^{}=-0.76\pm0.08$\, \cite{pdg}.} taking into account the two allowed regions differing by the sign of Re$(a,b)$.
The distributions of the points in relation to these parameters are exhibited in Fig.\,\,\ref{Bvsab}.
In the computation of $\cal B$ for this figure, we have devised functions $f_{a,b,c,d}^{}$ for Im$(a,b,c,d)$ which interpolate linearly between their relativistic (rel) and heavy-baryon (HB) expressions:
$f_k^{}={\rm Im}\,k_{\rm rel\,}^{} (1-y_k^{}) + {\rm Im}\,k_{\rm HB\,}^{} y_k^{}$\, for \,$k=a,b,c,d$\, and \,$0\le y_k^{}\le1$.\,
Accordingly, with respect to the horizontal axes, the points in the top two plots of Fig.\,\ref{Bvsab} lie, respectively, within the intervals \,$2.85\le{\rm Im}\,a(0)/{\rm MeV}\le6.86$ and~\,$-1.84\le{\rm Im}\,b(0)/{\rm MeV}\le-0.54$,\, as quoted earlier. In the bottom two plots, the $|{\rm Re}(a,b)|$ ranges are somewhat larger than those implied by the $|{\rm Re}(a,b)|$ numbers in Table\,\,\ref{smresults} because we have now extracted Re$(a,b)$ from the aforementioned \,$\Sigma^+\to p\gamma$\, data at 2$\sigma$.
Consequently, the benchmark points have a branching-fraction span of \,$1.2\times10^{-8}\le{\cal B}\le10.2\times10^{-8}$,\, which is somewhat wider than the previous prediction, Eq.\,(\ref{B2005}).
In these four plots, we have added horizontal red lines to mark the 2$\sigma$ upper-limit of the LHCb result~\cite{Aaij:2017ddf}.

\begin{figure}[!t] \bigskip
\includegraphics[height=63mm]{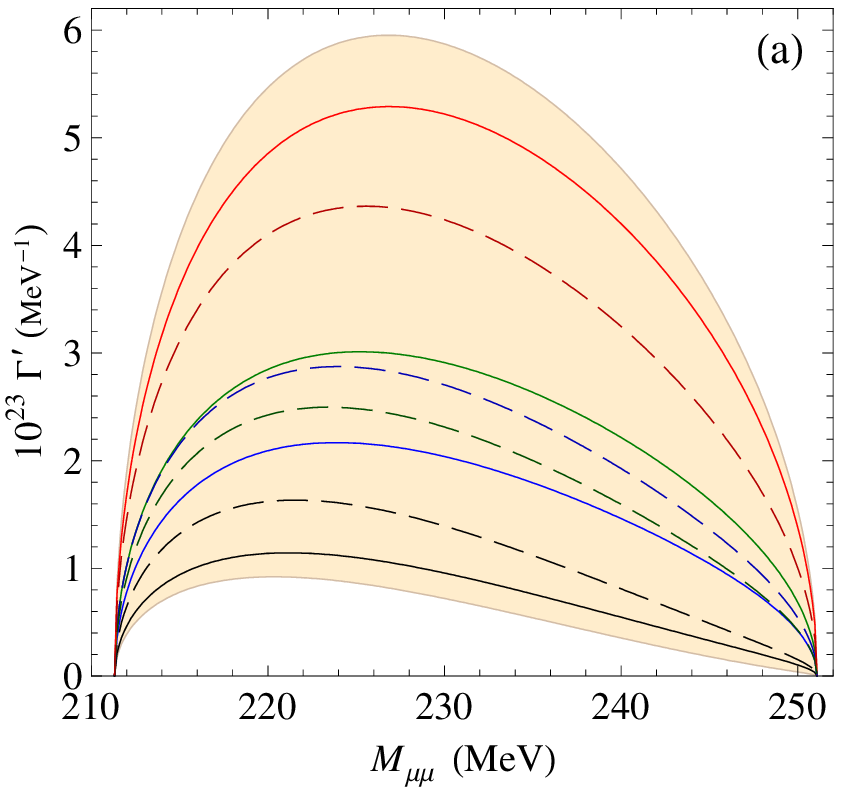} ~ ~ ~
\includegraphics[height=63mm]{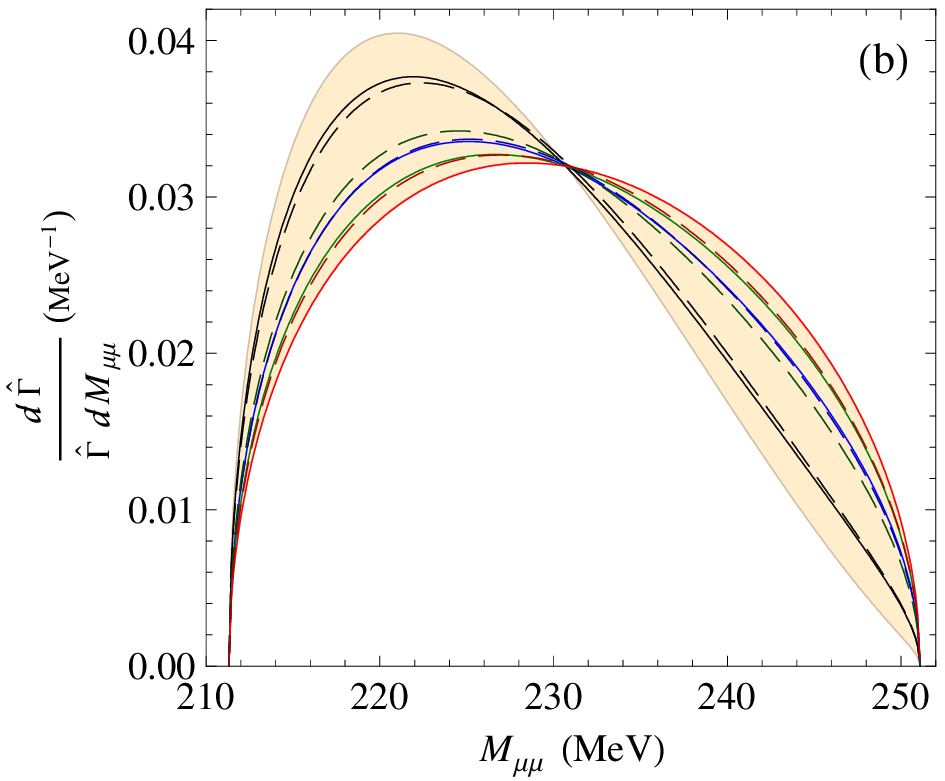} \vspace{-5pt}
\caption{(a) The dimuon invariant-mass distribution, \,$\Gamma'=d\Gamma(\Sigma^+\to p\mu^+\mu^-)/dq^2$\, versus \,$M_{\mu\mu}=\sqrt{q^2}$,\, calculated in the SM with Im$(a,b,c,d)$ formulas derived in relativistic (solid curves) or heavy-baryon (dashed curves) $\chi$PT.
From bottom to top, the black, blue, green, and red solid [dashed] curves correspond to
\,Re$(a,b)/{\rm MeV}=(13.3,-6.0),(-13.3,6.0),(6.0,-13.3),(-6.0,13.3)$
$[(11.0,-7.4),(-11.0,7.4)$, $(7.4,-11.0),(-7.4,11.0)]$, respectively.
The light-orange (shaded) region enveloping the curves corresponds to the parameter space represented by the benchmark points in Fig.\,\ref{Bvsab}.
(b) The related differential rate \,$\big(d\hat\Gamma/dM_{\mu\mu}\big)/\hat\Gamma=2\Gamma'M_{\mu\mu}/\hat\Gamma$\, normalized by  \,$\hat\Gamma=\Gamma(\Sigma^+\to p\mu^+\mu^-)$. \label{G'}} \vspace{-2ex}
\end{figure}

In Fig.\,\,\ref{G'}(a), we display the differential rate calculated in the SM as a function of the dimuon invariant mass, $M_{\mu\mu}$.
The light-orange (shaded) region depicts the range of the prediction from the parameter space represented by the benchmark points in Fig.\,\,\ref{Bvsab}.
To illustrate the prediction more specifically, we have drawn the solid [dashed] curves corresponding to the first [last] 4 sets of Re$(a,b)$ listed in Table\,\,\ref{smresults}.
In Fig.\,\,\ref{G'}(b) the related differential rate \,$\big(d\hat\Gamma/dM_{\mu\mu}\big)/\hat\Gamma=2\Gamma'M_{\mu\mu}/\hat\Gamma$,\, normalized by the total rate \,$\hat\Gamma\equiv\Gamma(\Sigma^+\to p\mu^+\mu^-)$,\, offers a complementary picture of the $M_{\mu\mu}$ distribution.
The $M_{\mu\mu}$ spectrum was also measured by LHCb~\cite{Aaij:2017ddf}, but more data are needed to test the prediction clearly.

\begin{figure}[h] \bigskip
\includegraphics[width=7cm]{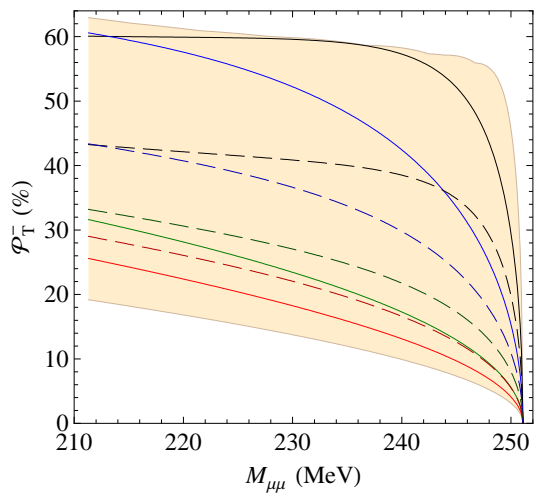} \vspace{-3pt}
\caption{The $\mu^-$ transverse-polarization asymmetry ${\cal P}_{\rm T}^-$ in \,$\Sigma^+\to p\mu^+\mu^-$\, versus $M_{\mu\mu}$ in the SM.
The light-orange (shaded) area represents the predicted range of ${\cal P}_{\rm T}^-$ and envelops curves corresponding to those  in Fig.\,\ref{G'} with the same curve styles. \label{PTsm}}
\end{figure}

As for the other observables of interest, we find that the forward-backward asymmetry ${\cal A}_{\rm FB}$ and the polarization asymmetries ${\cal P}_{\rm L,N}^-$ are tiny in the SM, below $10^{-4}$.
In light of Eqs.\,\,(\ref{AFB}), (\ref{PLNT-}), and\,\,(\ref{abcdefk}), this is attributable to the fact that all of these asymmetries involve no more than one factor from the LD component and therefore in the SM are proportional to at least one power of the product \,$\lambda_t^{~}y_{7A}^{}$,\, which is about \,$1.9\times10^{-6}$\, in size and comes from the SD amplitude.
For this reason, we do not show their graphs and only quote their integrated counterparts in Table\,\,\ref{smresults}.
In contrast, ${\cal P}_{\rm T}^-$ and $\tilde P_{\rm T}^-$ can be large, reaching up to roughly 60\%, because they each contain interference terms between two LD contributions.
This is illustrated in the last column of Table\,\,\ref{smresults} and in Fig.\,\,\ref{PTsm}, where the light-orange (shaded) region indicates the ${\cal P}_{\rm T}^-$ range and encloses solid and dashed curves corresponding to those in Fig.\,\,\ref{G'} with the same curve styles.
Concerning the $\mu^+$ polarization asymmetries, it is simple to see from Eqs.\,\,(\ref{PLNT+}) and\,\,(\ref{abcdefk}) that the SM predicts \,${\cal P}_{\rm L,N}^+={\cal P}_{\rm L,N}^-$\, and \,${\cal P}_{\rm T}^+\simeq{\cal P}_{\rm T}^-$.\,
This also applies to the integrated asymmetries $\tilde P_{\rm L,N,T}^\pm$.

The smallness of ${\cal A}_{\rm FB}$ and ${\cal P}_{\rm L,N}^\pm$ within the SM implies that they can serve as places to look for signals of physics beyond it.
Unambiguous measurements of any one of these quantities at the percent level or higher would be strong evidence for NP.
In the next section, we explore some scenarios which may bring about such signals.

\section{New physics contributions\label{bsm}}

As explained in the last section, in the SM the asymmetries ${\cal A}_{\rm FB}$ and ${\cal P}_{\rm L,N}^\pm$, as well as their integrated partners, are suppressed by at least one power of the coefficients in the decay amplitude which are not dominated by LD physics, namely $\tilde{\textsc e}$, $\tilde{\textsc f}$, and $\tilde{\textsc k}$, the other ones ($\tilde{\textsc g}$, $\tilde{\textsc h}$, and $\tilde{\textsc j}$) vanishing.
If new interactions beyond the SM exist, one or more of these coefficients could receive contributions which give rise to appreciable enlargement of some of the asymmetries and perhaps influence the rate as well.

To entertain this possibility, without dwelling on the specifics of an underlying model, for simplicity we focus on a couple of scenarios in which NP significantly affects only one of the coefficients which enter all of the asymmetries: $\tilde{\textsc e}$ and $\tilde{\textsc f}$.
Thus, we alter their SM formulas in Eq.\,(\ref{abcdefk}) to
\begin{align} \label{ef}
\tilde{\textsc e} & \,=\, \frac{G_{\rm F}^{}}{\sqrt2}\,\lambda_t^{~}y_{7A}^{}\,+\,g_{\textsc e\,}^{}e^{i\phi_{\textsc e}^{}} \,, ~~~~ ~~~~
\tilde{\textsc f} \,=\, \frac{D-F}{\sqrt2}\, G_{\rm F\,}^{}\lambda_t^{~} y_{7A}^{} \,+\, g_{\textsc f\,}^{} e^{i\phi_{\textsc f}^{}} \,,
\end{align}
where $g_{{\textsc e},{\textsc f}}^{}$ and $\phi_{{\textsc e},{\textsc f}}^{}$ stand for the magnitudes and phases of the NP contributions.

To investigate their numerical impact, for definiteness we fix the parameters of the SM amplitude to those of one of the benchmark points yielding the branching fraction \,${\cal B}^{\textsc{sm}}=2.0\times10^{-8}$, which also predicts the integrated asymmetry \,$\tilde P_{\rm T}^{-,\textsc{sm}}=54\%$,\, the other ones, $\tilde A_{\rm FB}^{\textsc{sm}}$ and $\tilde P_{\rm L,N}^{-,\textsc{sm}}$, being below $10^{-4}$ in size.
To restrain the NP terms, we impose \,${\cal B}\big(\Sigma^+\to p\mu^+\mu^-\big)<5.8\times10^{-8}$,\, which is the 2$\sigma$ upper-bound of the LHCb's finding~\cite{Aaij:2017ddf}.
This then translates into \,$g_{\textsc e}^{}<8.4\times10^{-8}{\rm\,GeV}^{-2}$\, if \,$g_{\textsc f}^{}=0$\, and \,$g_{\textsc f}^{}<1.2\times10^{-7}{\rm\,GeV}^{-2}$\, if \,$g_{\textsc e}^{}=0$.\,
In each of these scenarios, to explore the implications for the rate and asymmetries we consider a couple of examples: (I) $g_{\textsc e}^{}=7\times10^{-9}{\rm\,GeV}^{-2}$\, and \,$7\times10^{-8}{\rm\,GeV}^{-2}$\, if \,$g_{\textsc f}^{}=0$\, and (II) $g_{\textsc f}^{}=1\times10^{-8}{\rm\,GeV}^{-2}$\, and \,$1\times10^{-7}{\rm\,GeV}^{-2}$\, if \,$g_{\textsc e}^{}=0$.\,
In addition, we let $\phi_{{\textsc e},{\textsc f}}^{}$ run from 0 to 2$\pi$.
It turns out that the variation of these phases is unimportant for the rate but can have dramatic effects on the asymmetries.

In scenario I, where the NP enters via $\tilde{\textsc e}$ alone, the first example gives \,${\cal B}=2.0\times10^{-8}$,\, almost identical to its SM value, and $\tilde P_{\rm T}^-$ which fluctuates roughly around its SM prediction and between 48\% and 58\% depending on $\phi_{\textsc e}^{}$,\, as can be viewed in the top left graph of Fig.\,\ref{i,ii}.
Also in this instance $\tilde A_{\rm FB}$ and $\tilde P_{\rm L,N}^-$ are merely under 1\% in size.
In the second example, with $g_{\textsc e}^{}$ being ten times greater, we see more striking discrepancies from SM expectations: ${\cal B}=4.6\times10^{-8}$,\, more than twice its SM value, $\tilde P_{\rm T}^-$ varies between 1\% and 46\%, and $|\tilde P_{\rm L}^-|$ can reach about 3\%, whereas $|\tilde A_{\rm FB}|$ and $|\tilde P_{\rm N}^-|$ are still less than 1\%.
How these asymmetries deviate from their SM predictions and oscillate with $\phi_{\textsc e}^{}$ is displayed in the top right graph of Fig.\,\ref{i,ii}.

In scenario II, only the coefficient $\tilde{\textsc f}$ has the NP term.
In the first example, \,${\cal B}=2.0\times10^{-8}$\, and $\tilde P_{\rm T}^-$ fluctuates between 49\% and 57\%, similarly to the corresponding situation in scenario I.
On the other hand, in this case $\tilde P_{\rm L}^-$ can be fairly large, up to ${\cal O}(\pm10\%)$, and $\tilde A_{\rm FB}$ and $\tilde P_{\rm N}^-$ can reach a few percent, as the bottom left graph in Fig.\,\ref{i,ii} reveals.
In the second example, we get \,${\cal B}=4.7\times10^{-8}$\, and $\tilde P_{\rm T}^-$ varies between 7\% and 39\%, which are not very different from their counterparts in scenario I.
Nevertheless, as the bottom right graph in Fig.\,\ref{i,ii} shows, the other asymmetries enjoy the most enhancement compared to the preceding instances and can go up to as much as a few tens percent in size.

Extra information may be gained from the $\mu^+$ polarization asymmetries, which are not displayed here.
Since \,$\tilde{\textsc g}=\tilde{\textsc h}=\tilde{\textsc j}=0$\, in both scenarios and only $\tilde{\textsc e}$ or $\tilde{\textsc f}$ has a NP term, \,$\tilde P_{\rm L}^+=\tilde P_{\rm L}^-$\, and $\tilde P_{\rm T}^+$ is approximately out of phase with $\tilde P_{\rm T}^-$ in all these examples, while \,$\tilde P_{\rm N}^+=\tilde P_{\rm N}^-$ $\big(\tilde P_{\rm N}^+\simeq\tilde P_{\rm N}^-\big)$\, in scenario I (II).

\begin{figure}[t]
\includegraphics[height=53mm]{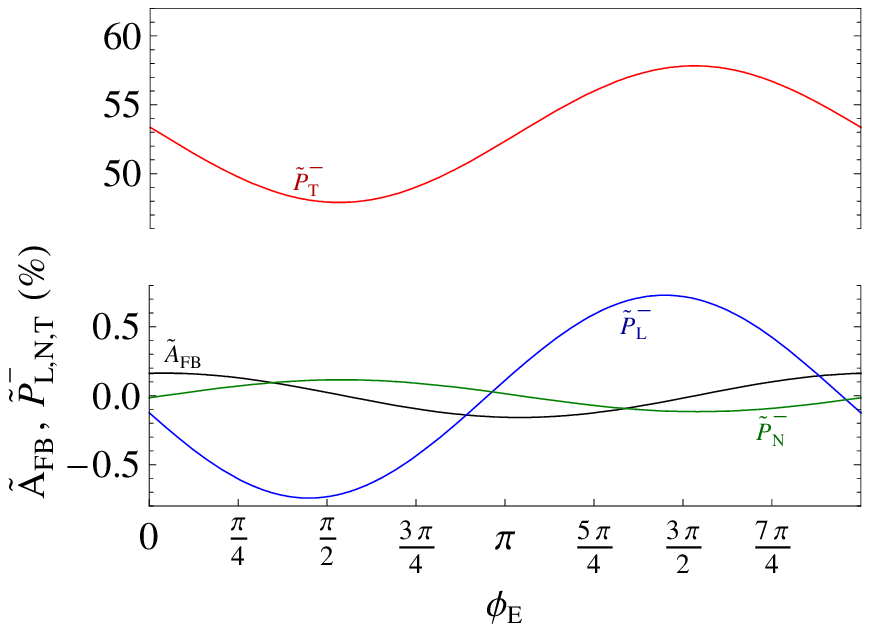} ~ ~ ~
\includegraphics[height=53mm]{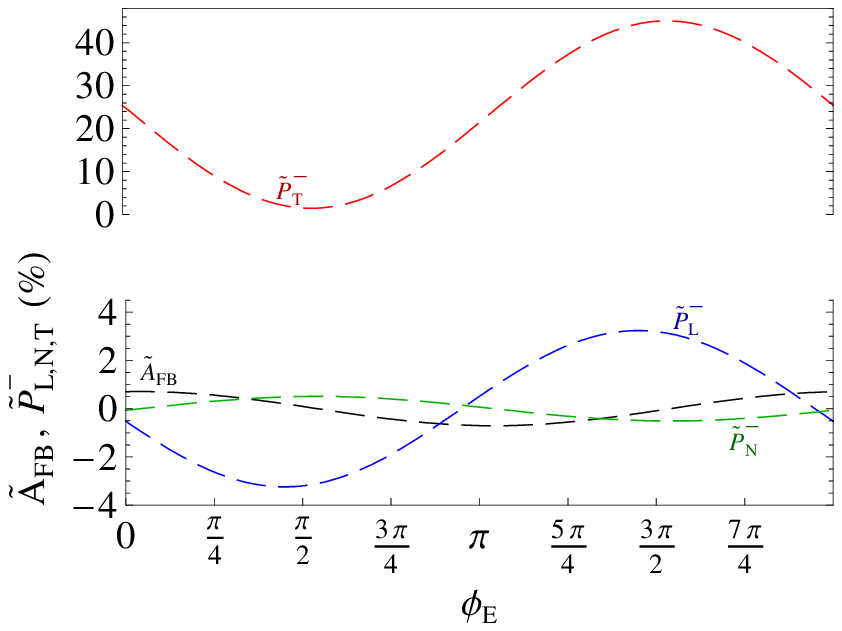}\vspace{7pt}\\
\includegraphics[height=53mm]{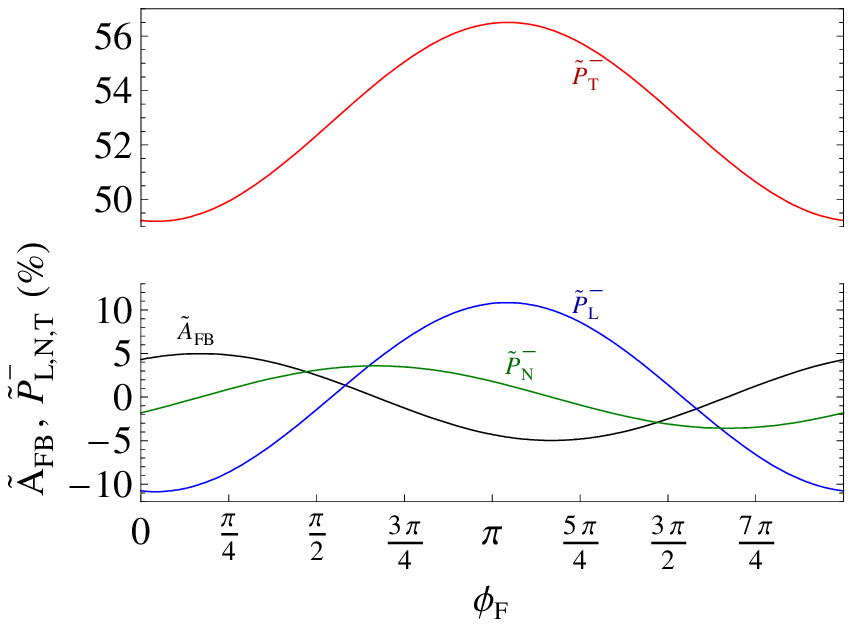} ~ ~
\includegraphics[height=53mm]{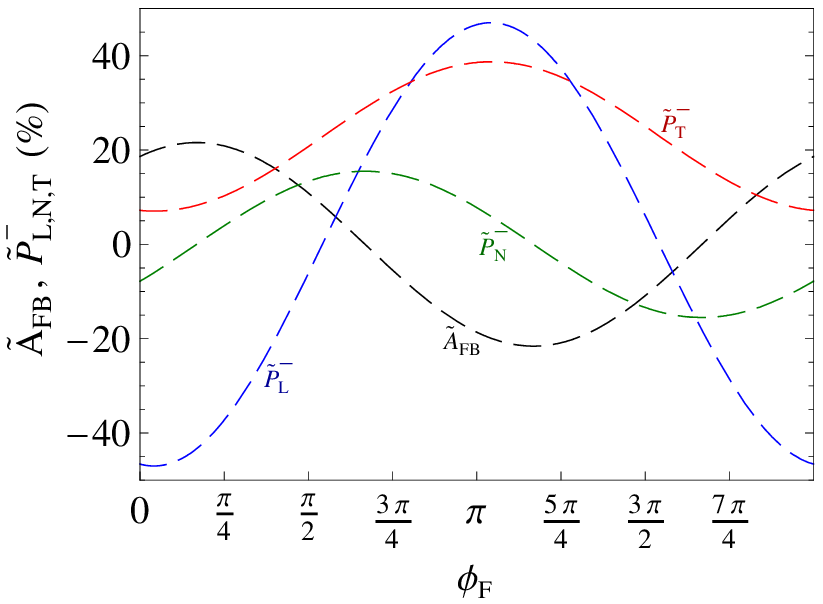}\vspace{-5pt}
\caption{The integrated asymmetries $\tilde A_{\rm FB}$ and $\tilde P_{\rm L,N,T}^-$ of the muon in \,$\Sigma^+\to p\mu^+\mu^-$\, versus the phases $\phi_{{\textsc e},{\textsc f}}^{}$ of the NP contributions to the coefficients $\tilde{\textsc e}$ (top plots) and $\tilde{\textsc f}$ (bottom plots), respectively, in the decay amplitude.
For the top plots, only $\tilde{\textsc e}$ has the NP term with magnitude \,$g_{\textsc e}^{}=7\times10^{-9}{\rm\,GeV}^{-2}$ (left) and \,$7\times10^{-8}{\rm\,GeV}^{-2}$ (right).
For the bottom plots, only $\tilde{\textsc f}$ has the NP term with magnitude \,$g_{\textsc f}^{}=1\times10^{-8}{\rm\,GeV}^{-2}$ (left) and \,$1\times10^{-7}{\rm\,GeV}^{-2}$ (right). \label{i,ii}}
\end{figure}

Finally, we remark that at the quark level the simplest effective interactions that could bring about the above NP contributions are described by \,${\cal L}_{\rm NP}\supset-\overline{d}\gamma^\kappa\big(\hat g_V^{}+\gamma_5^{}\hat g_A^{}\big)s\,
\overline{\mu}\gamma_\kappa^{}\gamma_5^{}\mu + {\rm H.c.}$\,
where $\hat g_V^{}$ and $\hat g_A^{}$ are coupling constants characterizing the underlying heavy physics.
However, strict restraints from the data on kaon decays \,$K\to\pi\mu^+\mu^-$\, and \,$K_{L,S}\to\mu^+\mu^-$ \cite{pdg,Mescia:2006jd} would prevent $\hat g_{V,A}^{}$, respectively, from being sufficiently large to generate the $g_{\textsc e}^{}$ and $g_{\textsc f}^{}$ terms of the desired size.
One might also think of attempting to evade the kaon restrictions by including other quark operators, such as \,${\cal L}_{\rm NP}\supset-\hat g_P^{}\, \overline{d}\gamma_5^{}s\,\overline{\mu}\gamma_5^{}\mu + {\rm H.c.}$\, which could help render the \,$K_{L,S}\to\mu^+\mu^-$\, constraints no longer applicable.
However, this would be achievable only with some degree of fine tuning among the different operators.
This indicates that if NP signals are detected in the asymmetries examined here, the underlying interactions may need to be described by more than a simple combination of the operators.

\section{Comments on \boldmath$\Sigma^+\to pe^+e^-$\label{s2pee}}

There is less empirical information available on this process than the muon mode.
The only existing measurement is \,$\Gamma\big(\Sigma^+\to pe^+e^-\big)/\Gamma\big(\Sigma^+\to p\pi^0\big)=(1.5\pm0.9)\times10^{-5}$\, \cite{Ang:1969hg}, which can be translated into  \,${\cal B}\big(\Sigma^+\to pe^+e^-\big)=(7.7\pm4.6)\times10^{-6}$\, \cite{He:2005yn}.\footnote{The PDG bound \,${\cal B}\big(\Sigma^+\to pe^+e^-\big)<7\times10^{-6}$\, \cite{pdg} is for the presence of neutral currents in this mode and not for its total branching fraction \cite{Ang:1969hg}.\medskip}
More data on this decay may be coming in the near future, as it will likely be one of the subjects of upcoming LHCb experiments on rare \,$s\to d$\, transitions~\cite{Bediaga:2018lhg}.

Theoretically, the dielectron invariant-mass distribution has already been studied within the SM in Ref.\,\cite{He:2005yn}, which predicted \,$9.1\times10^{-6}\le{\cal B}(\Sigma^+\to pe^+e^-)_{\rm SM}^{}\le10.1\times10^{-6}$,\, compatible with the preceding number.
We note that the relative uncertainty in this case is smaller than its muon counterpart.
This is due to the predominance in this mode of the $\Sigma^+$ undergoing nearly on-shell radiative weak decay and producing a Dalitz pair \cite{Dalitz:1951aj,Kroll:1955zu}.
The corresponding branching fraction can be written as
\begin{equation} \label{dalitz}
{\cal B}\big(\Sigma^+\to pe^+e^-\big)_{\rm Dalitz} \;=\;
\frac{2\alpha(0)}{3\pi} \Bigg[ {\rm ln}\,\frac{2\big(m_\Sigma^{}-m_p\big)}{m_e^{}}-\frac{13}{6} \Bigg]
{\cal B}\big(\Sigma^+\to p\gamma\big) \,,
\end{equation}
derived from Eq.\,(\ref{dG/ds}), with $\mu$ replaced by $e$, including only the SM LD parts.\footnote{To arrive at (\ref{dalitz}) involves taking some appropriate limits, such as setting the form factors $(a,b)$ to their \,$q^2=0$\, values. The result in (\ref{dalitz}) proportional to $|a(0)|^2$ is in line with the findings in the literature on electromagnetic Dalitz decays \cite{Kroll:1955zu}. For the term in (\ref{dalitz}) proportional to $|b(0)|^2$, the number \,$-$13/6\, roughly approximates the original one, $-$5/3, which amounts to decreasing the $|b(0)|^2$ term by about 10\%.}
For the benchmarks considered in section \ref{smpredict}, which yield
\,$8.4\times10^{-6}\le{\cal B}(\Sigma^+\to pe^+e^-)_{\rm SM}^{}\le11.0\times10^{-6}$,\, somewhat broader than the previous prediction quoted above, ${\cal B}(\Sigma^+\to pe^+e^-)_{\rm Dalitz}^{}$ accounts for 88\% to 99\% of the total,
${\cal B}(\Sigma^+\to pe^+e^-)_{\rm SM}^{}$.

Although the formulas obtained in section \ref{observables} are applicable to \,$\Sigma^+\to pe^+e^-$,\, not all of the observables proposed therein can be experimentally probed in the latter's case.
In particular, since the electron does not decay, measuring the $e^\pm$ polarization asymmetries are no longer feasible in \,$\Sigma^+\to pe^+e^-$.\,

Lastly, in light of the tentative hints of lepton-flavor universality (LFU) violation recently detected in a number of rare $b$-meson decays~\cite{Bediaga:2018lhg}, it is of interest to look for clues of LFU violation in other processes.
However, the possibility of testing LFU in \,$\Sigma^+\to p\ell^+\ell^-$ $(\ell=e,\mu)$ is curtailed by the substantial difference in phase space between the two modes and also obscured by the LD contributions which are not measurable in the radiative decay but which have relatively greater impact on the muon mode.
This is analogous to what occurs in the rare kaon decays \,$K_L\to\ell^+\ell^-$ \cite{Valencia:1997xe} and \,$K\to\pi\ell^+\ell^-$ \cite{Littenberg:1993qv}.

\section{Conclusions\label{concl}}

Motivated by the evidence for \,$\Sigma^+\to p\mu^+\mu^-$\, recently reported by LHCb and in anticipation of more data on this rare decay which they will collect in the near future, we have revisited our earlier SM calculation to present up-to-date predictions for the rate and the dimuon invariant-mass distribution.
Furthermore, we constructed a series of muon asymmetries for this mode which offer extra means to investigate it.
Given that within the SM this process is completely dominated by long-distance physics and thus subject to sizable uncertainties, we presented ranges of predictions for all these observables, taking into account the uncertainties as well as known constraints.
Similarly to our past estimates, our results for the branching fraction and dimuon invariant-mass distribution agree with the recent LHCb measurements which still have limited statistics.
Much improved data are needed to probe the predictions more stringently.

We demonstrated that some of the asymmetries are highly suppressed in the SM, which on the upside makes them potentially good avenues for new-physics searches.
We entertained the possibility of NP influencing a couple of the terms in the decay amplitude which are not dominated by LD physics and  differ in Dirac structure.
We illustrated how the different NP contributions can produce different effects on the aforementioned observables.
Our examples suggest that for the asymmetries expected to be negligible in the SM clear measurements at the percent level or higher would constitute a compelling hint of NP presence.

\acknowledgements

X.G.H thanks the hospitality of Shanxi Normal University during the initial stages of this work.
This research was supported in part by the MOE Academic Excellence Program (Grant No. 105R891505) and NCTS of ROC.
The work of X.G.H. was supported in part by the MOST of ROC (Grant No. MOST104-2112-M-002-015-MY3 and
106-2112-M-002-003-MY3), in part by the Key Laboratory for Particle Physics, Astrophysics and Cosmology,
Ministry of Education, and Shanghai Key Laboratory for Particle Physics and Cosmology
(Grant No. 15DZ2272100), and in part by the NSFC (Grant Nos. 11575111 and 11735010) of PRC.
The work of G.V. was supported in part by the Australian Research Council.
We would like thank Jeremy Dalseno and Francesco Dettori for comments and information on experimental matters.

\appendix

\section{Additional formulas\label{app}}

In the rest frame of the dimuon system, Eqs.\,\,(\ref{ft}) and (\ref{pc}) imply
\begin{align}
t & \,=\, m_\Sigma^2 + m_\mu^2 - 2E_\Sigma^{}E_\mu^{} +
2|\bm{p}_\Sigma^{}|\, |\bm{p}_\mu|\, c_\theta^{} \,, &
|\bm{p}_\mu^{}| & \,=\, \frac{\beta}{2}\sqrt{q^2} \,, &
\\
E_\Sigma^{} & \,=\, \frac{{\texttt M}_-^{} {\texttt M}_+^{}+q^2}{2\sqrt{q^2}} \,, ~~~ ~~~ ~~~
E_\mu^{} \,=\, \frac{\sqrt{q^2}}{2} \,, &
|\bm{p}_\Sigma^{}| & \,=\, \sqrt{\frac{\hat m_-^2\hat m_+^2}{4 q^2}} \,, &
\end{align}
and hence \,$f=\tfrac{1}{4}\big(\beta^2 c_\theta^2-1\big)\lambda$\, in Eq.\,(\ref{ft}).
From Eqs.\,\,(\ref{d^2G/ds/dt}) and (\ref{|M|^2}), we can then arrive at
\begin{equation}
\frac{d^2\Gamma(\Sigma^+\to p\mu^+\mu^-)}{dq^2\, dc_\theta^{}} \,=\,
\frac{\beta\sqrt{\bar\lambda}}{2}\, \frac{d^2\Gamma(\Sigma^+\to p\mu^+\mu^-)}{dq^2\,dt}
\,=\, {\cal F}_0^{} + {\cal F}_1^{}\, c_\theta^{} + {\cal F}_2^{}\, c_\theta^2 \,,
\end{equation}
where ${\cal F}_{0,1,2}^{}$ are independent of $\theta$ and given by
\begin{align}
{\cal F}_0^{} \,= &~\, \frac{\beta q^2 \sqrt{\bar\lambda}}{64 \pi^{3\,} m_\Sigma^3} \Bigg\{
\Bigg[ \frac{\big(2-\beta^2\big){\texttt M}_+^2+q^2}{4} |\tilde{\textsc a}|^2
+ \frac{4m_\mu^2+{\texttt M}_+^2+q^2}{4 q^2} |\tilde{\textsc c}|^2
+ \frac{3-\beta^2}{2}\, {\texttt M}_+^{}\, {\rm Re}(\tilde{\textsc a}{}^*\tilde{\textsc c})
\Bigg] \hat m_-^2
\nonumber \\ & \hspace{10ex}
+ \Bigg[ \frac{\big(2-\beta^2\big){\texttt M}_-^2+q^2}{4} |\tilde{\textsc b}|^2
+ \frac{4m_\mu^2+{\texttt M}_-^2+q^2}{4 q^2} |\tilde{\textsc d}|^2 - \frac{3-\beta^2}{2}\,
{\texttt M}_-^{}\, {\rm Re}(\tilde{\textsc b}{}^*\tilde{\textsc d}) \Bigg] \hat m_+^2
\nonumber \\ & \hspace{10ex}
+ \Bigg( \beta^2 \hat m_-^2 + \frac{{\texttt M}_-^2-\beta^2q^2}{2 q^2}\, m_+^2 \Bigg)
\frac{|\tilde{\textsc e}|^2}{2} + \Bigg( \beta^2 m_+^2 +
\frac{{\texttt M}_+^2-\beta^2q^2}{2q^2}\,m_-^2 \Bigg) \frac{|\tilde{\textsc f}|^2}{2} \Bigg\}
\nonumber \\ &
+ \frac{\beta \sqrt{\bar\lambda}}{256 \pi^{3\,} m_\Sigma^3}  \begin{array}[t]{l} \Big\{ \Big[
\big(\beta^2 |\tilde{\textsc g}|^2+|\tilde{\textsc j}|^2\big) q^2
+ 4 m_\mu^{}\, {\texttt M}_-^{}\, {\rm Re}(\tilde{\textsc e}{}^*\tilde{\textsc j}) \Big] \hat m_+^2
\vspace{3pt} \\ ~
+ \Big[ \big(\beta^2 |\tilde{\textsc h}|^2+|\tilde{\textsc k}|^2\big) q^2
- 4 m_\mu^{}\, {\texttt M}_+^{}\, {\rm Re}(\tilde{\textsc f}{}^*\tilde{\textsc k}) \Big] \hat m_-^2
\Big\} \,, \end{array}
\nonumber \\ \vphantom{\int^{\int_\int^\int}}
{\cal F}_1^{} \,= &~\, \frac{\beta^2 \bar\lambda}{64\pi^{3\,}m_\Sigma^3}\, {\rm Re} \!
\begin{array}[t]{l} \Big\{ \big[ {\texttt M}_+^{} \tilde{\textsc a}{}^* \tilde{\textsc f}
- {\texttt M}_-^{} \tilde{\textsc b}{}^* \tilde{\textsc e}
- \big( \tilde{\textsc a}{}^*\tilde{\textsc g}+\tilde{\textsc b}{}^*\tilde{\textsc h}\big) m_\mu
+ \tilde{\textsc c}{}^*\tilde{\textsc f} + {\textsc d}{}^*\tilde{\textsc e} \big] q^2
\vspace{3pt} \\ ~
- \big( {\texttt M}_+^{} \tilde{\textsc c}{}^* \tilde{\textsc g}
- {\texttt M}_-^{} \tilde{\textsc d}{}^* \tilde{\textsc h} \big) m_\mu \Big\} \,, \end{array}
\nonumber \\ \label{F2} \vphantom{\int^{\int_\int^\int}}
{\cal F}_2^{} \,= &~\, \frac{\beta^3\bar\lambda^{3/2}}{256\pi^{3\,} m_\Sigma^3} \Big[
\big(|\tilde{\textsc a}|^2+|\tilde{\textsc b}|^2\big) q^2 - |\tilde{\textsc c}|^2
- |\tilde{\textsc d}|^2 - |\tilde{\textsc e}|^2 - |\tilde{\textsc f}|^2 \Big] \,.
\end{align}
With Eqs.\,\,(\ref{dG/ds}) and (\ref{F2}), it is straightforward to check that
\,$\Gamma'=2{\cal F}_0^{}+\tfrac{2}{3} {\cal F}_2^{}$.\,
Numerically, ${\cal F}_2^{}/3$ turns out to be small compared to ${\cal F}_0^{}$
and consequently \,$\Gamma'\simeq2{\cal F}_0^{}$.\,


\begin{thebibliography}{0}

\bibitem{Park:2005ek}
  H.~Park {\it et al.}  [HyperCP Collaboration],
  %``Evidence for the decay Sigma+ $\to$ p mu+ mu-,''
  Phys.\ Rev.\ Lett.\  {\bf 94}, 021801 (2005)  [arXiv:hep-ex/0501014].
  %%CITATION = HEP-EX 0501014;%%

\bibitem{Aaij:2017ddf}
  R.~Aaij {\it et al.} [LHCb Collaboration],
  %``Evidence for the rare decay $\Sigma^+ \to p \mu^+ \mu^-$,''
  Phys.\ Rev.\ Lett.\  {\bf 120}, 221803 (2018)  [arXiv:1712.08606 [hep-ex]].
  %%CITATION = doi:10.1103/PhysRevLett.120.221803;%%

\bibitem{He:2005yn}
  X.G.~He, J.~Tandean, and G.~Valencia,
  %``The decay Sigma+ $\to$ p l+ l- within the standard model,''
Phys.\ Rev.\ D {\bf 72}, 074003 (2005)  [hep-ph/0506067].
  %%CITATION = doi:10.1103/PhysRevD.72.074003;%%

\bibitem{Santos:2018zbz}
  D.M.~Santos [LHCb Collaboration],
  %``A strange program for LHCb,''
  EPJ Web Conf.\  {\bf 179}, 01013 (2018).
  %%CITATION = doi:10.1051/epjconf/201817901013;%%

\bibitem{He:2005we}
  X.G.~He, J.~Tandean, and G.~Valencia,
  %``Implications of a new particle from the HyperCP data on Sigma+ $\to$ p l+l-,''
  Phys.\ Lett.\ B {\bf 631}, 100 (2005)  [hep-ph/0509041].
  %%CITATION = doi:10.1016/j.physletb.2005.10.005;%%

\bibitem{hypercpx}
%\bibitem{Deshpande:2005mb}
  N.G.~Deshpande, G.~Eilam, and J.~Jiang,
  %``On the possibility of a new boson X0 (214-MeV) in Sigma+ $\to$ p mu+ mu-,''
  Phys.\ Lett.\ B {\bf 632}, 212 (2006); %  [arXiv:hep-ph/0509081];
  %%CITATION = HEP-PH 0509081;%%
%\bibitem{Geng:2005ra}
  C.Q.~Geng and Y.K.~Hsiao,  %``Constraints on the new particle in Sigma+ $\to$ p mu+ mu-,''
  Phys.\ Lett.\ B {\bf 632}, 215 (2006); % [arXiv:hep-ph/0509175];
  %%CITATION = HEP-PH 0509175;%%
%\bibitem{Gorbunov:2005nu}
  D.S.~Gorbunov and V.A.~Rubakov,  %``On sgoldstino interpretation of HyperCP events,''
  Phys.\ Rev.\ D {\bf 73}, 035002 (2006); % [arXiv:hep-ph/0509147];
  %%CITATION = HEP-PH 0509147;%%
%\bibitem{Demidov:2006pt}
  S.V.~Demidov and D.S.~Gorbunov,  %``More about sgoldstino interpretation of HyperCP events,''
  JETP Lett.\  {\bf 84}, 479 (2007); % [arXiv:hep-ph/0610066];
  %%CITATION = JTPLA,84,479;%%
%\bibitem{He:2006uu}
  X.G.~He, J.~Tandean, and G.~Valencia,  %``Light Higgs production in hyperon decay,''
  Phys.\ Rev.\  D {\bf 74}, 115015 (2006); % [arXiv:hep-ph/0610274];
  %%CITATION = PHRVA,D74,115015;%%
%\bibitem{He:2006fr}
%  X.G.~He, J.~Tandean, and G.~Valencia,
%``Does the HyperCP Evidence for the Decay Sigma --> p mu+ mu- Indicate a Light Pseudoscalar Higgs Boson?,''
  Phys.\ Rev.\ Lett.\  {\bf 98}, 081802 (2007); % [hep-ph/0610362];
  %%CITATION = doi:10.1103/PhysRevLett.98.081802;%%
%\bibitem{He:2008zw}
%  X.G.~He, J.~Tandean, and G.~Valencia,
  %``Rare Decays with a Light CP-Odd Higgs Boson in the NMSSM,''
  JHEP {\bf 0806}, 002 (2008); % [arXiv:0803.4330];
  %%CITATION = JHEPA,0806,002;%%
%\bibitem{Chen:2006xja}
  C.H.~Chen and C.Q.~Geng,  %``Implications of the HyperCP data on B and tau decays,''
  Phys.\ Lett.\  B {\bf 645}, 189 (2007); % [arXiv:hep-ph/0612142];
  %%CITATION = PHLTA,B645,189;%%
%\bibitem{Zhu:2006zv}
  S.h.~Zhu,  %``Unique Higgs boson signature at colliders,''
  hep-ph/0611270;
  %%CITATION = HEP-PH/0611270;%%
%\bibitem{Mangano:2007gi}
  M.L.~Mangano and P.~Nason,
  %``Radiative quarkonium decays and the NMSSM Higgs interpretation of the
  %hyperCP Sigma+ --> p mu+ mu- events,''
  Mod.\ Phys.\ Lett.\  A {\bf 22}, 1373 (2007); % [arXiv:0704.1719];
  %%CITATION = MPLAE,A22,1373;%%
%\bibitem{Chen:2007uv}
  C.H.~Chen, C.Q.~Geng, and C.W.~Kao,  %``U-boson and the HyperCP exotic events,''
  Phys.\ Lett.\  B {\bf 663}, 400 (2008); % [arXiv:0708.0937];
  %%CITATION = PHLTA,B663,400;%%
%\bibitem{Tatischeff:2007dz}
  B.~Tatischeff and E.~Tomasi-Gustafsson,
  %``Search for Low Mass Exotic mesonic structures. Part I: experimental %results,''
  Phys.\ Part.\ Nucl.\ Lett.\  {\bf 5}, 363 (2008); % [arXiv:0710.1796];
  %%CITATION = 00438,5,363;%%
%\bibitem{Xiangdong:2007vv}
  X.~Gao, C.S.~Li, Z.~Li, and H.~Zhang,
%``Contributions from SUSY-FCNC couplings to the interpretation of the HyperCP events for the decay Sigma+ ---> p mu+ mu-,''
  Eur.\ Phys.\ J.\ C {\bf 55}, 317 (2008); %  [arXiv:0712.0257 [hep-ph]].
  %%CITATION = doi:10.1140/epjc/s10052-008-0580-z;%%
%\bibitem{Heng:2008rc}
  Z.~Heng, R.J.~Oakes, W.~Wang, Z.~Xiong, and J.~M.~Yang,
  %``B meson Dileptonic Decays in NMSSM with a Light CP-odd Higgs Boson,''
  Phys.\ Rev.\  D {\bf 77}, 095012 (2008); % [arXiv:0801.1169];
  %%CITATION = PHRVA,D77,095012;%%
%\bibitem{Hou:2008di}
  G.W.S.~Hou,
  %``Search for TeV Scale Physics in Heavy Flavour Decays,''
  Eur.\ Phys.\ J.\ C {\bf 59}, 521 (2009); %  [arXiv:0808.1932 [hep-ex]];
  %%CITATION = doi:10.1140/epjc/s10052-008-0688-1;%%
%\bibitem{Chang:2008np}
  Q.~Chang and Y.D.~Yang,
  %``Rare decay \pi^{0}\to e^+ e^- as a sensitive probe of light CP-odd Higgs in NMSSM,''
  Phys.\ Lett.\ B {\bf 676}, 88 (2009); % [arXiv:0808.2933];
  %%CITATION = doi:10.1016/j.physletb.2009.04.081;%%
%\bibitem{Das:2008at}
  P.K.~Das,
%``Implication of the HyperCP boson $X^0$ (214 MeV) in the flavour changing neutral current processes,''
  Phys.\ Rev.\  D {\bf 80}, 034017 (2009); % [arXiv:0809.0498];
  %%CITATION = PHRVA,D80,034017;%%
%\bibitem{Cheung:2008rh}
  K.~Cheung and T.J.~Hou,
    %``Light Pseudoscalar Higgs boson in Neutralino Decays in the Next-to-Minimal
  %Supersymmetric Standard Model,''
  Phys.\ Lett.\  B {\bf 674}, 54 (2009); % [arXiv:0809.1122];
  %%CITATION = PHLTA,B674,54;%%
%\bibitem{McKeen:2008gd}
  D.~McKeen,  %``Constraining Light Bosons with Radiative $\Upsilon(1S)$ Decays,''
  Phys.\ Rev.\  D {\bf 79}, 015007 (2009); % [arXiv:0809.4787];
  %%CITATION = PHRVA,D79,015007;%%
%\bibitem{Cao:2008rc}
  J.~Cao and J.M.~Yang,  %``Anomaly of Zbb coupling revisited in MSSM and NMSSM,''
  JHEP {\bf 0812}, 006 (2008); % [arXiv:0810.0751];
  %%CITATION = JHEPA,0812,006;%%
%\bibitem{Pospelov:2008zw}
  M.~Pospelov,  %``Secluded U(1) below the weak scale,''
  Phys.\ Rev.\ D {\bf 80}, 095002 (2009); % [arXiv:0811.1030];
  %%CITATION = doi:10.1103/PhysRevD.80.095002;%%
%\bibitem{Reece:2009un}
M.~Reece and L.T.~Wang,  %``Searching for the light dark gauge boson in GeV-scale experiments,''
  JHEP {\bf 0907}, 051 (2009); % [arXiv:0904.1743];
  %%CITATION = JHEPA,0907,051;%%
%\bibitem{Hooper:2009gm}
  D.~Hooper and T.M.P.~Tait,
%``Neutralinos in an extension of the minimal supersymmetric standard model as the source of the PAMELA positron excess,''
  Phys.\ Rev.\ D {\bf 80}, 055028 (2009); % [arXiv:0906.0362 [hep-ph]];
  %%CITATION = doi:10.1103/PhysRevD.80.055028;%%
%\bibitem{Tian:2009ar}
  L.J.~Tian, Y.L.~Jin, and Y.~Jiang,
%``Yangian description for decays and possible explanation of X in the decay K0(L) ---> pi0 pi0 X,''
  Phys.\ Lett.\ B {\bf 686}, 207 (2010); %  [arXiv:0910.0588];
  %%CITATION = doi:10.1016/j.physletb.2010.02.040;%%
%\bibitem{Ellwanger:2009dp}
U.~Ellwanger, C.~Hugonie, and A.M.~Teixeira,
%``The Next-to-Minimal Supersymmetric Standard Model,''
  Phys.\ Rept.\  {\bf 496}, 1 (2010); % [arXiv:0910.1785];
  %%CITATION = doi:10.1016/j.physrep.2010.07.001;%%
%\bibitem{Oh:2009fm}
  S.~Oh and J.~Tandean,  %``Rare B Decays with a HyperCP Particle of Spin One,''
  JHEP {\bf 1001}, 022 (2010); % [arXiv:0910.2969];
  %%CITATION = doi:10.1007/JHEP01(2010)022;%%
%\bibitem{Rashed:2010jp}
  A.~Rashed, M.~Duraisamy, and A.~Datta,
  %``Probing light pseudoscalar, axial vector states through $\eta_{b} \to\tau^{+}\tau^{-}$,''
  Phys.\ Rev.\ D {\bf 82}, 054031 (2010); %  [arXiv:1004.5419 [hep-ph]];
  %%CITATION = doi:10.1103/PhysRevD.82.054031;%%
%\bibitem{Andreas:2010ms}
  S.~Andreas, O.~Lebedev, S.~Ramos-Sanchez, and A.~Ringwald,
  %``Constraints on a very light CP-odd Higgs of the NMSSM and other axion-like particles,''
  JHEP {\bf 1008}, 003 (2010); %  [arXiv:1005.3978 [hep-ph]];
  %%CITATION = doi:10.1007/JHEP08(2010)003;%%
%\bibitem{Almarashi:2011hj}
  M.M.~Almarashi and S.~Moretti,
  %``Muon Signals of Very Light CP-odd Higgs states of the NMSSM at the LHC,''
  Phys.\ Rev.\ D {\bf 83}, 035023 (2011); % [arXiv:1101.1137];
  %%CITATION = doi:10.1103/PhysRevD.83.035023;%%
%\bibitem{Tang:2012zk}
  L.~Tang, H.W.~Ke, and X.Q.~Li,
  %``Study on the effects of the light CP-odd Higgs via the leptonic decays of pseudoscalar mesons,''
  Commun.\ Theor.\ Phys.\  {\bf 58}, 732 (2012); % [arXiv:1205.4474];
  %%CITATION = doi:10.1088/0253-6102/58/5/18;%%
%\bibitem{Ke:2012wa}
  H.W.~Ke, X.H.~Yuan, X.Q.~Li, Z.T.~Wei, and Y.X.~Zhang,
  %``$\Sigma_{b}\to\Sigma_c$ and $\Omega_b\to\Omega_c$ weak decays in the light-front quark model,''
  Phys.\ Rev.\ D {\bf 86}, 114005 (2012). % [arXiv:1207.3477];
  %%CITATION = doi:10.1103/PhysRevD.86.114005;%%

\bibitem{exp} %{Kaplan:2007nn}
  D.M.~Kaplan,
  %``A New experiment to study hyperon CP violation and the charmonium system,''
  Int.\ J.\ Mod.\ Phys.\ A {\bf 22}, 5958 (2007); %  [arXiv:0707.1543 [hep-ex]].
  %%CITATION = doi:10.1142/S0217751X07039158;%%
%\bibitem{Love:2008aa}
  W.~Love {\it et al.} [CLEO Collaboration],
  %``Search for Very Light CP-Odd Higgs Boson in Radiative Decays of Upsilon(S-1),''
  Phys.\ Rev.\ Lett.\  {\bf 101}, 151802 (2008); %  [arXiv:0807.1427 [hep-ex]].
  %%CITATION = doi:10.1103/PhysRevLett.101.151802;%%
%\bibitem{Tung:2008gd}
  Y.C.~Tung {\it et al.} [E391a Collaboration],
  %``Search for a light pseudoscalar particle in the decay K0(L) ---> pi0 pi0 X,''
  Phys.\ Rev.\ Lett.\  {\bf 102}, 051802 (2009); % [arXiv:0810.4222];
  %%CITATION = doi:10.1103/PhysRevLett.102.051802;%%
%\bibitem{Park:2010zze}
  H.K.~Park,
  %``Search for the HyperCP event at the LHCb experiment,''
  JHEP {\bf 1010}, 052 (2010);
  %%CITATION = doi:10.1007/JHEP10(2010)052;%%
%\bibitem{Aubert:2009cp}
  B.~Aubert {\it et al.} [BaBar Collaboration],
  %``Search for Dimuon Decays of a Light Scalar Boson in Radiative Transitions Upsilon ---> gamma A0,''
  Phys.\ Rev.\ Lett.\  {\bf 103}, 081803 (2009); % [arXiv:0905.4539 [hep-ex]].
  %%CITATION = doi:10.1103/PhysRevLett.103.081803;%%
%\bibitem{Hyun:2010an}
  H.J.~Hyun {\it et al.} [Belle Collaboration],
%``Search for a Low Mass Particle Decaying into mu^+mu^- in B^0->K^{*0}X and B^0->rho^0X at Belle,''
  Phys.\ Rev.\ Lett.\  {\bf 105}, 091801 (2010); % [arXiv:1005.1450];
  %%CITATION = doi:10.1103/PhysRevLett.105.091801;%%
%\bibitem{Abouzaid:2011mi}
  E.~Abouzaid {\it et al.} [KTeV Collaboration],
%``Search for the Rare Decays $K_L \to \pi^0\pi^0\mu^+\mu^-$ and
%$K_L \to \pi^0\pi^0 X^0\to \pi^0\pi^0\mu^+\mu^-$,''
  Phys.\ Rev.\ Lett.\  {\bf 107}, 201803 (2011); % [arXiv:1105.4800];
  %%CITATION = doi:10.1103/PhysRevLett.107.201803;%%
%\bibitem{Ablikim:2011es}
  M.~Ablikim {\it et al.} [BESIII Collaboration],
  %``Search for a light Higgs-like boson $A^0$ in $J/\psi$ radiative decays,''
  Phys.\ Rev.\ D {\bf 85}, 092012 (2012); % [arXiv:1111.2112];
  %%CITATION = doi:10.1103/PhysRevD.85.092012;%%
%\bibitem{Komatsubara:2012pn}
  T.K.~Komatsubara,
  %``Experiments with K-Meson Decays,''
  Prog.\ Part.\ Nucl.\ Phys.\  {\bf 67}, 995 (2012); % [arXiv:1203.6437];
  %%CITATION = doi:10.1016/j.ppnp.2012.04.001;%%
%\bibitem{Echenard:2012hq}
  B.~Echenard,
  %``Search for Light New Physics at B Factories,''
  Adv.\ High Energy Phys.\  {\bf 2012}, 514014 (2012); %  [arXiv:1209.1143 [hep-ex]];
  %%CITATION = doi:10.1155/2012/514014;%%
%\bibitem{Li:2016tlt}
  H.B.~Li,
  %``Prospects for rare and forbidden hyperon decays at BESIII,''
  Front.\ Phys.\ (Beijing) {\bf 12}, no. 5, 121301 (2017). % [arXiv:1612.01775].
  %%CITATION = doi:10.1007/s11467-017-0691-9;%%

\bibitem{Buchalla:1995vs}
G.~Buchalla, A.J.~Buras, and M.E.~Lautenbacher,
%``Weak Decays Beyond Leading Logarithms,''
Rev.\ Mod.\ Phys.\  {\bf 68}, 1125 (1996) [arXiv:hep-ph/9512380].
%%CITATION = HEP-PH 9512380;%%

\bibitem{Bergstrom:1987wr}
I.V. Lyagin and E.K. Ginzburg, %``On $\Sigma^+\to pe^+e^-$ and $\Sigma^+\to p\mu^+\mu^-$ decays,''
Sov. Phys. JETP {\bf 14}, 653 (1962);
L.~Bergstrom, R.~Safadi, and P.~Singer,
%``Phenomenology Of Sigma+ to P Lepton+ Lepton- And The Structure Of The%Weak Nonleptonic Hamiltonian,''
  Z.\ Phys.\ C {\bf 37}, 281 (1988).
  %%CITATION = ZEPYA,C37,281;%%

\bibitem{pdg} % {Tanabashi:2018oca}
  M.~Tanabashi {\it et al.} [Particle Data Group],
  %``Review of Particle Physics,''
  Phys.\ Rev.\ D {\bf 98}, no. 3, 030001 (2018). % doi:10.1103/PhysRevD.98.030001
  %%CITATION = doi:10.1103/PhysRevD.98.030001;%%

\bibitem{b2sll} %{Hewett:1995dk}
  J.L.~Hewett,
  %``Tau polarization asymmetry in B ---> X(s) tau+ tau-,''
  Phys.\ Rev.\ D {\bf 53}, 4964 (1996)  [hep-ph/9506289];
  %%CITATION = doi:10.1103/PhysRevD.53.4964;%%
%\bibitem{Kruger:1996cv}
  F.~Kruger and L.M.~Sehgal,
  %``Lepton polarization in the decays b ---> X(s) mu+ mu- and B ---> X(s) tau+ tau-,''
  Phys.\ Lett.\ B {\bf 380}, 199 (1996)  [hep-ph/9603237];
  %%CITATION = doi:10.1016/0370-2693(96)00413-3;%%
%\bibitem{Guetta:1997fw}
  D.~Guetta and E.~Nardi,
  %``Searching for new physics in rare B ---> tau decays,''
  Phys.\ Rev.\ D {\bf 58}, 012001 (1998)  [hep-ph/9707371];
  %%CITATION = doi:10.1103/PhysRevD.58.012001;%%
%\bibitem{Fukae:1998qy}
  S.~Fukae, C.S.~Kim, T.~Morozumi, and T.~Yoshikawa,
  %``A Model independent analysis of the rare B decay B ---> X(s) lepton+ lepton-,''
  Phys.\ Rev.\ D {\bf 59}, 074013 (1999)  [hep-ph/9807254];
  %%CITATION = doi:10.1103/PhysRevD.59.074013;%%
%\bibitem{Fukae:1999ww}
  S.~Fukae, C.S.~Kim, and T.~Yoshikawa,
%``A Systematic analysis of the lepton polarization asymmetries in the rare B decay, B--->X(s)tau+tau-,''
  Phys.\ Rev.\ D {\bf 61}, 074015 (2000)  [hep-ph/9908229];
  %%CITATION = doi:10.1103/PhysRevD.61.074015;%%
%\bibitem{Bensalem:2002ni}
  W.~Bensalem, D.~London, N.~Sinha, and R.~Sinha,
  %``Lepton polarization and forward backward asymmetries in b ---> s tau+ tau-,''
  Phys.\ Rev.\ D {\bf 67}, 034007 (2003)  [hep-ph/0209228].
  %%CITATION = doi:10.1103/PhysRevD.67.034007;%%

\bibitem{Lb2Lll} %{Chen:2001zc}
  C.H.~Chen and C.Q.~Geng,
  %``Baryonic rare decays of Lambda(b) ---> Lambda lepton+ lepton-,''
  Phys.\ Rev.\ D {\bf 64}, 074001 (2001)  [hep-ph/0106193];
  %%CITATION = doi:10.1103/PhysRevD.64.074001;%%
%\bibitem{Aliev:2002ww}
  T.M.~Aliev, A.~Ozpineci, and M.~Savci,
  %``Exclusive $\Lambda_b \to \Lambda \ell^{+} \ell^{-}$ decay beyond standard model,''
  Nucl.\ Phys.\ B {\bf 649}, 168 (2003)  [hep-ph/0202120];
  %%CITATION = doi:10.1016/S0550-3213(02)00964-1;%%
%\bibitem{Giri:2005mt}
  A.K.~Giri and R.~Mohanta,
%``Study of FCNC mediated Z boson effect in the semileptonic rare baryonic decays
%Lambda(b) ---> Lambda l+ l-,''
  Eur.\ Phys.\ J.\ C {\bf 45}, 151 (2006)  [hep-ph/0510171].
  %%CITATION = doi:10.1140/epjc/s2005-02407-6;%%

\bibitem{Savage:1990km}
  M.J.~Savage and M.B.~Wise,
  %``Polarization in K+ ---> pi+ mu+ mu-,''
  Phys.\ Lett.\ B {\bf 250}, 151 (1990);
  %%CITATION = doi:10.1016/0370-2693(90)91170-G;%%
%\bibitem{Agrawal:1991sh}
  P.~Agrawal, J.N.~Ng, G.~Belanger, and C.Q.~Geng,
  %``A Study of T violation in K+ ---> pi+ mu+ mu- decays,''
  Phys.\ Rev.\ D {\bf 45}, 2383 (1992).
  %%CITATION = doi:10.1103/PhysRevD.45.2383;%%

\bibitem{Mescia:2006jd}
  F.~Mescia, C.~Smith, and S.~Trine,
%``K(L)--->pi0 e+ e- and K(L)--->pi0 mu+ mu-: A Binary star on the stage of flavor physics,''
  JHEP {\bf 0608}, 088 (2006)  [hep-ph/0606081];
  %%CITATION = doi:10.1088/1126-6708/2006/08/088;%%
%\bibitem{DAmbrosio:2017klp}
  G.~D'Ambrosio and T.~Kitahara,
  %``Direct $CP$ Violation in $K \to \mu^+ \mu^-$,''
  Phys.\ Rev.\ Lett.\  {\bf 119}, no. 20, 201802 (2017)  [arXiv:1707.06999 [hep-ph]].
  %%CITATION = doi:10.1103/PhysRevLett.119.201802;%%

\bibitem{Davier:2017zfy}
  M.~Davier, A.~Hoecker, B.~Malaescu, and Z.~Zhang,
 %``Reevaluation of the hadronic vacuum polarisation contributions to the Standard Model
 %predictions of the muon $g-2$ and ${\alpha (m_Z^2)}$ using newest hadronic cross-section data,''
  Eur.\ Phys.\ J.\ C {\bf 77}, no. 12, 827 (2017)  [arXiv:1706.09436 [hep-ph]].
  %%CITATION = doi:10.1140/epjc/s10052-017-5161-6;%%

\bibitem{Bijnens:1985kj}
  J.~Bijnens, H.~Sonoda, and M.B.~Wise,
  %``On the Validity of Chiral Perturbation Theory for Weak Hyperon Decays,''
  Nucl.\ Phys.\ B {\bf 261}, 185 (1985).
  %%CITATION = doi:10.1016/0550-3213(85)90569-3;%%

\bibitem{Ang:1969hg}
  G.~Ang {\it et al.},
  %``Radiative sigma-plus-minus decays and search for neutral currents,''
  Z.\ Phys.\  {\bf 228}, 151 (1969).
  %%CITATION = doi:10.1007/BF01397536;%%

\bibitem{Bediaga:2018lhg}
  I.~Bediaga {\it et al.} [LHCb Collaboration],
%``Physics case for an LHCb Upgrade II - Opportunities in flavour physics, and beyond, in the HL-LHC era,''
  arXiv:1808.08865.
  %%CITATION = ARXIV:1808.08865;%%

\bibitem{Dalitz:1951aj}
  R.H.~Dalitz,
  %``On an alternative decay process for the neutral pi-meson, Letters to the Editor,''
  Proc.\ Phys.\ Soc.\ A {\bf 64}, 667 (1951).
  %%CITATION = doi:10.1088/0370-1298/64/7/115;%%

\bibitem{Kroll:1955zu}
  N.M.~Kroll and W.~Wada,
  %``Internal pair production associated with the emission of high-energy gamma rays,''
  Phys.\ Rev.\  {\bf 98}, 1355 (1955);
  %%CITATION = doi:10.1103/PhysRev.98.1355;%%
%\bibitem{Pilkuhn:1971uk}
  H.~Pilkuhn,
  %``Dalitz pair rates in the decays of omega, phi, k*, x and sigma0,''
  Nucl.\ Phys.\ B {\bf 29}, 462 (1971).
  %%CITATION = doi:10.1016/0550-3213(71)90035-6;%%

\bibitem{Valencia:1997xe}
  G.~Valencia,
  %``Long distance contribution to K(L) ---> lepton+ lepton-,''
  Nucl.\ Phys.\ B {\bf 517}, 339 (1998)  [hep-ph/9711377].
  %%CITATION = doi:10.1016/S0550-3213(98)00116-3;%%

\bibitem{Littenberg:1993qv}
  L.~Littenberg and G.~Valencia,
  %``Rare and radiative kaon decays,''
  Ann.\ Rev.\ Nucl.\ Part.\ Sci.\  {\bf 43}, 729 (1993)  [hep-ph/9303225].
  %%CITATION = doi:10.1146/annurev.ns.43.120193.003501;%%

\end{thebibliography}
\end{document}